
 \magnification=1200
\vfill
\line{\hfil UB-ECM-PF-94/16}
\line{\hfil November 1994}
\vskip 2truecm
\font\gran=cmbx12
{ \centerline{\gran Renormalization and the Equivalence Theorem:
On-shell Scheme }}
\vskip 1truecm
\bigskip
\centerline{D. Espriu\footnote{$^*$}{E-mail:
espriu@ebubecm1.ecm.ub.es;
espriu@greta.ecm.ub.es} ~and~ J. Matias\footnote{$^\dagger$}{E-mail:
matias@ebubecm1.ecm.ub.es}}
\bigskip \centerline{\it D.E.C.M., Facultat de F\'\i sica and I.F.A.E.}
\centerline{\it Universitat de Barcelona}
\centerline{\it Diagonal, 647}
\centerline{\it E-08028 Barcelona}
\vskip 1.5truecm
\centerline{ABSTRACT}
 \bigskip

We perform an exhaustive analysis of the Equivalence Theorem both
in the minimal Standard Model and in an Effective Electroweak Chiral
Lagrangian up to ${\cal O}(p^4)$. We have considered the leading
corrections to the usual prescription consisting in just replacing
longitudinally polarized $W$ or $Z$ by the corresponding Goldstone
bosons. The corrections
appear through an overall
constant multiplying the Goldstone boson amplitude as well as through
additional
diagrams. By including them we can extend the domain of applicability
of the Equivalence Theorem, making it
suitable for precision tests of the symmetry breaking sector
of the Standard Model. The on-shell scheme has been used throughout.
When considering the Equivalence Theorem in an Effective
Chiral lagrangian we analyze its
domain
of applicability, as well as several side issues concerning gauge
fixing, Ward
identities, on-shell scheme and matching
conditions in the effective theory.
We have  analyzed in detail the processes $W^+ W^- \to W^+W^-$
and $W^+W^+\to W^+W^+$  to illustrate the points made.

\vfill
\eject

\line{\bf 1. Introduction\hfil}
\medskip\noindent
The Equivalence Theorem states that for any spontaneously broken
gauge theory, provided the energy transfer is large enough, one can
replace the longitudinal degrees of freedom of the massive vector
bosons by the appropriate Goldstone bosons and use them to compute
$S$-matrix elements.

Even though the Equivalence Theorem was proved originally [1-2]
in the context of the minimal Standard Model,
with a doublet of complex scalar fields, it has been
realized[3-5] that it should remain valid for other theories
exhibiting
an equivalent set of fields and symmetries, even for non renormalizable
ones.

This makes the Equivalence Theorem
potentially very useful in investigations
of the symmetry breaking sector of the Standard Model. Indeed, it has
become customary to describe such a
sector by a non-linear, non-renormalizable
effective theory, the Effective Chiral Lagrangian[6].
 The
Goldstone
bosons of the broken global symmetry $SU(2)_L\times SU(2)_R\to SU(2)_V$
are collected  in a matrix-valued dimensionless field $U(x)$.
The operators in the Effective Chiral Lagrangian are classified according
to the number of derivatives or gauge fields acting on $U(x)$.
 At low
energies only the first terms in the expansion are of interest as
the typical size of the expansion parameter in the minimal Standard Model
is $p^2/(4\pi v)^2$ or $p^2/M_H^2$, whichever is larger.
Since $v=250$ GeV,
if the Higgs is very heavy the expansion is clearly a very good one.
On the other hand, if Nature has ruled that
Higgs does not exist, we have
to appeal to Technicolor or composite models
to account for the breaking of
the global symmetry and the appeareance of the Goldstone bosons. The
coefficients of the Effective Chiral Lagrangian will then differ
from the values they take in the minimal Standard Model.
However, it would be unrealistic to expect dramatic
changes in their order of magnitude.  Thus it is not difficult
to convince oneself that the
contribution from the ${\cal O}(p^4)$ operators  is all that it will
be possible to measure in the near future (the contribution
from the ${\cal O}(p^2)$ operators is universal and carries no
information on whatever underlying physics gives the {\it Z} and {\it W}
a mass).

Formally the non-linear, non-renormalizable Effective Chiral Lagrangian
is very similar to the long distance effective description of
strong interactions in terms of pions and kaons, the strong
chiral lagrangian[7]. Much of what over the years has been
learnt from the interactions of pions and kaons can then be easily
taken over to the weak interaction case. For this purpose
the Equivalence Theorem is instrumental.

In recent times there has been a flurry of papers dealing in one
way or another with the Equivalence Theorem. The activity
has proceeded mostly along two directions. First, it has been
realized that the common textbook statement
of the Equivalence Theorem, namely
$$\eqalign{A(W_L W_L \to W_L \dots W_L)=&
 (-i)^{n} A(\omega \omega \to \omega \dots \omega )
+{\cal O}(M/\sqrt{s}),} \eqno(1.1)$$
where $\omega$ denotes the goldstone boson `eaten' by the
appropriate $W$ or $Z$ boson,
is not quite correct. For one thing,
there is an overall factor ${\cal C}^n$ on the r.h.s of (1.1) [8-10].
The origin and relevance of this factor we will
discuss in detail in the coming sections. Moreover
it is incorrect to discard all
the terms that in (1.1) are lumped together in
the `${\cal O}(M/\sqrt{s})$' bit. Except in
the crudest of approximations both terms
need to be kept. As we will show they both give subleading
corrections that do not necessarily vanish in the large $s$
limit and
that are needed if one wants to perform a detailed comparison with the
experimental results. It should be stated
right away that the Equivalence Theorem has
been mostly used in the context of the so-called
strongly interacting Higgs --- the limit in which
the quartic self coupling in the scalar sector of
the minimal Standard Model becomes large. In this
limit the dominant contributions are
correctly taken into account by (1.1). Yet, the
corrections to (1.1) are not negligible at all. This is the origin of
some misgivings that have been raised[11-13]
concerning the usefulness of the Equivalence Theorem.
After studying this issue in detail,
we hope
to convince the reader that
the Equivalence Theorem remains a powerful tool
to disentangle the scalar sector of the Standard Model.

Second, we all know that the minimal Standard Model need not
be the correct theory for the symmetry breaking sector. We know that the
Higgs
is particularly well hidden  in the Electroweak Theory[14]. It makes
a lot of sense to try and investigate possible departures
from the minimal Standard Model by setting bounds on
the ${\cal O}(p^4)$ coefficients of the Effective
Chiral Lagrangian. Scattering of longitudinal $W$'s and
$Z$'s are amongst  the clearest ways of doing this and the
Equivalence Theorem comes handy. As we have mentioned, it should
remain valid in effective, non-renormalizable theories. However,
these have a limited range of validity as we scale up the
energy (the upper bound being $4\pi v$ or $\Lambda$, $\Lambda$ being
the mass of the first resonance in the strongly interacting
scalar sector, whichever is smallest), so the energy
cannot be too large. It cannot be too small either
on account of the uncalculated pieces on the r.h.s of (1.1),
so the range of validity seems to be limited[13].
We will show
that taking properly into account the next to leading corrections to
the
Equivalence Theorem improves considerably the situation and allows for
practical applications with the required level of precision.

The accuracy reached in many experiments testing the
Electroweak
Theory is such that radiative corrections have necessarily to be taken
into account. Nowadays there seems to exist ample consensus in choosing
the on-shell scheme to carry out the renormalization program[15-16].
For this reason we have elected to work within this scheme
in our discussion of the Equivalence Theorem as is conceptually simple
and technically convenient.  To our knowledge this is the first
time that such an analysis is presented.

In deriving the above results we have been led to
a number of colateral issues.
We believe that some of the results  are interesting in their own
right and we have made an effort to collect them either in the main
body of the paper or in the appendices. Amongst these
we should mention: a discussion of the renormalization
in the longitudinal sector in the on-shell scheme,
Ward identities in the non-linear realization, modifications
to the on-shell scheme when the Higgs is not present, and
the full lagrangian expanded up to terms with four fields
in the non-linear variables. To keep the discussion simple we have
restricted ourselves to the charged sector. No conceptually new issues
appear in the neutral sector, but the $\gamma-Z$ mixing complicates the
analysis considerably.

Let us discuss briefly the way the paper is organized.
In section 2 we review some Ward identities in the
minimal Standard Model and discuss the formulation
of the Electroweak Theory in terms of the non-linear variables
suitable in the large Higgs mass limit. Section 3 is devoted
to the derivation of the Equivalence Theorem and the ${\cal C}$ factor in
the minimal Standard
Model. We then proceed to apply, in section 4, the previous results to the
analysis of the processes $W^+W^-\to W^+W^-$ and $W^+W^+\to W^+W^+$ in
the minimal Standard Model. The gauge-fixing
procedure, Ward identities and the extension of the Equivalence Theorem to
the
Effective Chiral Lagrangian is discussed in sections 5 and 6.
We have also included in section 5 a
discussion on the matching conditions  and how the results of
the minimal Standard Model are reproduced by a particular choice of the
${\cal O}(p^4)$ coefficients in the Effective Chiral Lagrangian.
In section  7 we apply the Equivalence Theorem
to the process $W^+W^-\to W^+W^-$ in
the Effective Chiral Lagrangian,
including a discussion on the domain of applicability.
Finally, our conclusions are summarized in section 8.

\bigskip

\line{\bf 2. Gauge Fixing and Ward Identities in
the Minimal Standard Model\hfil}
\medskip\noindent
We shall start our discussion
by outlining the derivation of some Ward identities
in the minimal Standard Model involving the
longitudinal sector of the theory.
We will first analyze the Standard Model in the usual linear
variables and move afterwards to the non-linear representation, more
suitable to deal with a strongly interacting scalar sector.
Throughout this paper we will work in the on-shell scheme renormalization
scheme and
we shall basically adhere to the conventions of [16] and [17-19],
except for a field redefinition in the scalar sector.

\bigskip
\line{\bf 2.1 Linear Realization\hfil}
\medskip\noindent
The minimal Standard Model lagrangian with a doublet of complex fields
$\Phi$ is
$$
{\cal L}_{SM}= D_{\mu} \Phi^{\dagger} D^{\mu} \Phi -
{\lambda } \vert\Phi\vert^4 + \mu^2 \vert\Phi\vert^2+
{\cal L}_{YM} +{\cal L}_{GF}+{\cal L}_{FP},
\eqno(2.1.1) $$
with $D_{\mu}=\partial_{\mu}+{1 \over 2} i g W_{\mu}^{i}
\sigma^{i}+{1 \over 2} i g^\prime B_{\mu}$. The Higgs doublet is
$$\Phi={1
\over \sqrt{2}} \left(\matrix {\omega^{2}+i\omega^{1} \cr
\sigma- i \omega^3}\right).\eqno(2.1.2)$$
In the notation of [16] $\omega^{2} \to \varphi^{1}$, $\omega^{1} \to
-\varphi^{2}$ and $\omega^{3} \to - \chi$.
The gauge fixing and Faddeev-Popov terms are
$${\cal L}_{GF} =-{1 \over 2} (2F^{+} F^{-} + F^{3} F^{3} + F^{0}
F^{0}), \qquad
{\cal L}_{FP} =\sum_{\alpha,\beta=i,0} {\bar c}^{\alpha}
{ \delta F^{\alpha} \over
\delta \theta^{\beta}} c^{\beta},\eqno(2.1.3)$$
with $i=1,2,3$ running over $SU(2)_L$ indices and $\alpha=0$ corresponding
to the $U(1)_Y$ part
$$\eqalign{
F^{\pm} &= {1 \over \sqrt{\xi_{1}^W}} \partial^{\mu}
W_{\mu}^{\pm} -  M_{W} \sqrt{\xi_{2}^{W}} \omega^{\pm}, \cr
F^{3} &={1 \over \sqrt{\xi_{1}^3}} \partial^{\mu} W_{\mu}^{3}
-M_{W} \sqrt{\xi_{2}^{3}}\, \omega^{3}, \cr
F^{0} &={1 \over \sqrt{\xi_{1}^B}} \partial_{\mu} B^{\mu}
+(M_{Z}^2-M_{W}^2)^{1 \over 2} \sqrt{\xi_{2}^{B}}\,
\omega^{3}.}\eqno(2.1.4) $$
The charged fields $W_\mu^\pm$ and $\omega^\pm$ are defined
by $W_\mu^\pm=(W^1_\mu\mp iW^2_\mu)/\sqrt{2}$ and
$\omega^\pm=(\omega^1\mp i\omega^2)/\sqrt{2}$.
In this work we will not consider
neutral fields and we will drop the indices `$W$' and `$B$',
being understood that unless stated otherwise we will be referring to
the charged fields. Notice that in the on-shell scheme $\xi_1$
and $\xi_2$, although equal classically, have to be kept different beyond
tree level because
they renormalize differently. Furthermore, the introduction of two gauge
parameters allows for the elimination of the divergences appearing in
mixed $W-\omega$ Green functions. This is discussed in Appendix A.

The starting point in our discussion is the generating functional of
connected Green functions
$$Z[J,\eta,{\bar \eta}]=\int {\cal D} X {\cal D} {\bar c}
{\cal D} c \exp \{ i \int d^4 x [ {\cal L}_{SM} + X^{\alpha}
J^{\alpha}+
{\bar c}^{\alpha} \eta^{\alpha} + {\bar \eta}^{\alpha} c^{\alpha}] \} .
\eqno(2.1.5)$$
($X^\alpha$ collectively denotes the fields in the theory, and
$\bar{c}$, $c$ and $\eta$, $\bar{\eta}$ are the ghosts and their sources,
respectively). A summation over space-time points as well as external
indices is understood. By differentiating twice and setting the sources
equal to
zero we obtain the bare propagators of the theory. In the longitudinal sector
we have gauge, Goldstone boson and mixed propagators. Their decomposition in
terms of invariant functions is
$$\eqalign{D_{\mu \nu}^{W^{+} W^{-}}(k)&=(-g_{\mu \nu} + {k_{\mu}
k_{\nu}
\over k^2})\Lambda^{W^{+} W^{-}}_{T}(k^2) - {k_{\mu} k_{\nu} \over
k^2} \Lambda^{W^{+} W^{-}}_{L}(k^2),\cr
D^{W^{+} \omega^{-}}_{\mu}(k)&=i k_{\mu} \Lambda^{W^{+} \pi^{-}}(k^2),
\cr
D^{\omega^{+} \omega^{-}}(k)&=\Lambda^{\pi^{+}
\pi^{-}}(k^2).}\eqno(2.1.6)$$
In terms of self-energies
$$\eqalign{
\Lambda^{W^{+} W^{-}}_{T}&={i \over k^2 - M_{0}^2 +
\Sigma_{T}}, \cr
\Lambda^{W^{+} W^{-}}_{L}&={i \xi_{1}^{0} \over k^2 -
\xi_{1}^{0} M_{0}^2 + \xi_{1}^{0} \Sigma_{L}},
\cr
\Lambda^{\omega^{+} \omega^{-}}&={i \over k^2 - \xi_{2}^{0}
M_0^2 + \Sigma_{\omega }},\cr
\Lambda^{W^+ \omega^-} &={i\xi_{1}^{0} \over k^{2} - M_{0}^{2}
\xi_{1}^{0}} \Sigma_{W\omega} {1 \over k^{2} -M^{2}
\xi_{2}^{0}}.}\eqno(2.1.7) $$
The last expression is valid at the one-loop level only.
We will restrict ourselves to this order as going beyond this
in the Standard Model is only of
academic interest at present.
Since we will be dealing exclusively with charged $W$'s, in order
not to unnecessarily clutter our formulae
we have
suppresed the indices in masses and self-energies whenever no
confussion is possible.
The relation between bare ($\Sigma$) and renormalized ($\hat{\Sigma}$)
self-energies is given in Appendix A. For the mixed propagator we have
only considered the one-loop expression in  (2.1.7) because this
is all
we will need in what follows. Since we work in a 't Hooft gauge there
is no tree level contribution to $\Lambda^{W^+\omega^-}$.
Furthermore note that with our conventions
$\Lambda^{W^+\omega^-}=-\Lambda^{\omega^+ W^-}=\Lambda^{W^-\omega^+}$
on account of the hermiticity of the effective action.

When the sources are set to zero the generating functional  (2.1.5)
is invariant under the BRS transformation ($\zeta^{2}=0$)
$$ \eqalign{
\delta W_{\mu}^i&=(-\delta^{ik}\partial_{\mu} +g
\epsilon^{ijk} W_{\mu}^{j}) c^{k} {\zeta}, \cr
\delta B_\mu &=-\partial_\mu c^0\zeta,\cr
\delta \omega^{i}&={g \over 2} ( \sigma \delta^{i k} +
\epsilon^{ijk} \omega^j)
 c^{k}\zeta -{g^\prime \over 2} (\sigma \delta^{i 3} -
\epsilon^{ij 3 }\omega^j) c^{0}\zeta, \cr
\delta\sigma &=-{g\over 2}\omega^i c^i\zeta +{g^\prime\over 2} \omega^3 c^0
\zeta,\cr
\delta \bar{c}^\alpha &=F^\alpha  \zeta, \cr
\delta c^i &=-{1 \over 2} \epsilon^{ijk}
c^{j} c^{k} \zeta,\cr
\delta c^0 &=0.\cr}
\eqno(2.1.8)$$
The fact that the gauge group is $SU(2)_L\times U(1)_Y$ makes
the BRS transformation somewhat involved. Using the invariance
of the action we arrive at
$$\langle 0\vert J^\gamma \delta X^\gamma +
\delta\bar{c}^\gamma \eta^\gamma+ \bar{\eta}^\gamma \delta c^\gamma
\vert 0\rangle_{J,\eta,\bar{\eta}}=0. \eqno(2.1.9)$$
A further derivation w.r.t. $\eta$ followed by the limit
$\eta=\bar{\eta}=0$ gives
$$\langle 0\vert F^\beta(X(y))\zeta\vert 0\rangle_J=\langle 0\vert \delta
\bar{c}^\beta (y)\vert
0\rangle_J =-i \langle 0\vert J^\gamma \delta X^\gamma \bar{c}^\beta (y)
\vert 0 \rangle_J. \eqno(2.1.10)$$
Acting now with $F^\alpha({\delta\over {i\delta J(x)}})$ on both sides of
(2.1.10), using that the gauge fixing is {\it linear} in the fields
and the fact that
$$ F^\alpha(\delta X(x))= ({\delta F^\alpha\over {\delta \theta^\sigma}}
c^\sigma)(x)\zeta,\eqno(2.1.11)$$
we get[16]
$$ F^\alpha({\delta\over {i\delta J(x)}}) F^\beta({\delta\over {i\delta
J(y)}}) Z[J]\vert_{J=0}= i\delta^{\alpha\beta}\delta(x-y) Z[0].
\eqno(2.1.12)$$

To be specific let us concentrate in the gauge fixing
condition for the charged fields, $F^{\pm}$. Eq. (2.1.12)
can be easily written in terms of propagators. Using (2.1.6)
$${k^2} \Lambda^{W^{+} W^{-}}_{L} - 2  M_{0}k^2 \sqrt{ \xi_{1
}^{0} \xi_{2}^{0}} \Lambda^{W^+ \omega^-}- \xi_{1}^{0}
\xi_{2}^{0} M_{0}^2 \Lambda^{\omega^{+} \omega^{-}}= i \xi_{1
}^{0}. \eqno(2.1.13) $$
We have used that $\Lambda^{W^+\omega^-}=-\Lambda^{\omega^+ W^-}$.
Now we substitute (2.1.7) into (2.1.13). If we work
at tree level we can
set $\xi_1^0=\xi_2^0=\xi$ and (2.1.13) is just an identity.
At the next order we have to keep track of the self-energies,
which are ${\cal O}(g^2)$ and of the difference between
$\xi_1^0$ and $\xi_2^0$ creeping in from the tree level
expression. Then one finally gets with one-loop precision
$$
k^{2} \xi_{1}^{0}\Sigma_{L}  -M_{0}^{2} \Sigma_\omega\xi_{2}^{0}
+ 2 M_{0} k^{2} \sqrt{ \xi_{1}^{0}
\xi_{2}^{0}} \Sigma_{W \omega}
=0 \eqno(2.1.14) $$
These Ward identities will be useful on two counts. On the one hand
they allow to express the mixed propagator and self-energy in terms of
the $W$ and $\omega$ propagators and self-energies. Moreover they
provide important relations between the different renormalization
constants in the longitudinal sector.
The relation between the bare and  renormalized expressions
is obtained through the use of the renormalization constants
described in Appendix A. We shall demand that the renormalized
gauge parameters are equal, i.e. $\xi_1=\xi_2$. Since
$Z_{\xi_1}\neq Z_{\xi_2}$ this requires $\xi_1^0\neq \xi_2^0$.
As a consequence
 there is a net counterterm for the self-energy
$\Sigma_{W\omega}(k^2)$
beyond tree level. (Recall that in t'Hooft gauges the mixed
$W-\omega$ piece cancels off at tree level between the gauge gauge
fixing term and the kinetic piece $(D_\mu\Phi)^\dagger D^\mu\Phi$
once the symmetry is broken,
$\langle \sigma\rangle=v$, and we shift
$\sigma\to v+\sigma$.)

In terms of renormalized quantities (2.1.14) reads
$$\eqalign{k^2 ( \hat{\Sigma}_{L} + 2 M \hat{\Sigma}_{W \omega} )
-M^{2}\hat{\Sigma}_{\omega} =& (k^2 - \xi M^2 )
( {k^2 \over \xi} (\delta Z_{W} - \delta Z_{\xi_{1}}
)
 -M^{2} (\delta Z_{\omega} + \delta Z_{M} + \delta
Z_{\xi_{2}} ) ).} \eqno(2.1.15)
$$
Notice that (2.1.15) provides us with some combinations of renormalization
constants which are necessarily finite, such as $Z_W Z_{\xi_1}^{-1}$
and $Z_{\omega} Z_M Z_{\xi_2}$.

We could have used ---in principle--- a
gauge condition other than (2.1.4). This
would have not affected the physical $S$-matrix elements, but would
have changed the formulation of the Ward identities needed  to
prove the Equivalence Theorem and, of course, reshuffled different
contributions amongst different diagrams. Some of the difficulties
encountered in other gauges
will we commented upon next.

\bigskip
\line{\bf 2.2. Non-linear Realization \hfil}
\medskip\noindent
If the Higgs mass is large, i.e. if the quartic coupling $\lambda$ is
large, the lagrangian (2.1.1) is not written in the most
convenient set of variables. Indeed, the quartic coupling
affects all four scalar fields $(\sigma, \omega^{i})$, thus involving
the Goldstone bosons of the $SU(2)_L\times SU(2)_R\to
SU(2)_V$ breaking which mix with the longitudinal
$W$'s and $Z$. On the other hand, since $M_H^2=
2\lambda v^2$, the internal exchange of the Higgs
boson is strongly suppressed. These two facts together lead to a
tremendous amount of cancellation between different diagrams, a
fact well know to any practitioner of the linear
sigma model[20].

It is far more convenient to rewrite the scalar sector
of the Standard Model using a non-linear realization.
The way to proceed is to introduce the matrix-valued
field $M(x)$
$$M(x)=\sqrt{2}(\matrix{\tilde{\Phi} \Phi}),\eqno(2.2.1)$$
where $\tilde{\Phi}$ is the hypercharge conjugated doublet.
We then perform the change of variables
$$M=\rho U\qquad U=\exp {i\over v}\pi^i\sigma^i,\eqno(2.2.2)$$
with $U\in SU(2)_L\times SU(2)_R/SU(2)_V$. The unitary
matrix $U$ collects the Goldstone bosons of the broken
global symmetry $\pi^i$. The fields $\pi^i$ are related
by a non-linear tranformation (involving the $\rho$ field)
to the $\omega^i$
used in the previous subsection.
Since $M$ transforms linearly,
so does $U$, but the Goldstone bosons themselves transform
non-linearly
$$U^\prime(x)= e^{{i\over 2} \alpha^{i}(x) \sigma^{i} }
 U(x) e^{- {i\over 2}
\alpha^{0}(x) \sigma^{3}},\eqno(2.2.3)$$
$$\eqalign{\delta
\pi^{i} =&{1 \over 2} \left( v (\alpha^{i} - \alpha^{0}
\delta^{i 3}) - \epsilon^{i j k} \alpha^{j} \pi^{k} + \alpha^{0} (
\pi^{2} \delta^{i 1 } - \pi^{1} \delta ^{i 2} ) \right)
 \cr
&+ {1 \over 6 v} \pi^{j} \pi^{l} \left( \alpha^{k} ( \delta^{l i}
\delta^{ k j} - \delta^{j l} \delta^{ k i} ) - \alpha^{0}
( \delta^{j 3} \delta^{l i} - \delta^{i 3} \delta^{j l} )
\right) + \dots }
\eqno (2.2.4)
$$
The field $\rho$ is inert under the gauge group. This is
in contradistinction to what happens to the $\sigma$ field
in a linear realization. The field $\rho$ gets a v.e.v.
when the symmetry is broken and, as usual, we shift
$\rho\to v+\rho$.
The steps required in going from the linear to the non-linear
realization have been discussed in detail in [19].

The lagrangian in the non-linear realization is
$$\eqalign{
{\cal L}_{SM}=&{1 \over 2} \partial_{\mu} \rho \partial^{\mu} \rho
-\rho {\lambda v} (v^2+{\mu^2\over \lambda})
-{1 \over 2} \rho^2 (\mu^2 + 3 v^2 \lambda)
- \lambda v \rho^3 - {1\over 4}\lambda\rho^4 \cr
&+{1 \over 4} (\rho+v)^2 {\rm Tr}D_{\mu} U^{\dag} D^{\mu} U +
{\cal L}_{GF}
+ {\cal L}_{FP}.}\eqno(2.2.5)$$
with $D_{\mu} U=\partial_{\mu} U + {1 \over 2} i g W_{\mu}^{i}
\sigma^{i} U(x) - {1 \over 2} i g^\prime B_{\mu} U(x) \sigma^{3}  $.
For the gauge-fixing part one could just use the transformed
of (2.1.4)
$$\eqalign{
{\cal L}_{GF}
\equiv &
 -{1\over {2}}\sum_{1=1,3}F^iF^i-{1\over {2}}F^0F^0 \cr
=&-{1 \over 2\xi^{W}}\sum_{i=1,3} (\partial^{\mu}
W_{\mu}^{i}
+ {i \over 4} g v \xi^{W} (\rho+v) {\rm Tr}\tau^{i} U)^2 \cr
&-{1 \over 2\xi^{B}}(\partial^{\mu} B_{\mu} -
{i \over {4 }} g^\prime v \xi^{B}(\rho +v){\rm Tr}\tau^{3} U)^2.\cr
} \eqno(2.2.6) $$
(We shall not distinguish here between $\xi_1$ and $\xi_2$
to keep our formulae manageable). Finally
$$\eqalign{
{\cal L}_{FP}&= \partial^\mu c^{0\dag} \partial_{\mu}
c^{0} +
 \partial^\mu c^{i\dag} \partial_{\mu} c^{i}
+g\, c^{i\dag} (\partial^{\mu} W_{\mu}^{k} \epsilon ^{ikj}
- {1 \over 8} g v \xi^{W}(\rho+v){\rm Tr} \tau^{i} \tau^{j} U) c^{j} \cr
&-{1 \over 8} g^{\prime 2} v \xi^{B} c^{0\dag}(\rho+v){\rm Tr} U
c^{0} +{1 \over 8} \sqrt{g g^\prime}
\,v \, (g^\prime \xi^{B}
c^{0\dag} c^{i}+g \xi^{W} c^{i\dag} c^{0})(\rho+v){\rm Tr}
\tau^{3} \tau^{i} U. }
\eqno(2.2.7)$$
In Landau gauge ($\xi=0$) ghosts decouple from
Goldstone bosons.

${\cal L}_{SM}+{\cal L}_{GF}+{\cal L}_{FP}$ is invariant
under BRS tranformations. The gauge fields
transform as in (2.1.8) and
$$ \eqalign{
\delta \rho &=0,\cr
\delta U &={i\over 2} g \sigma^i U c^i\zeta -{i\over 2} g^{\prime}
U \sigma^3 c^0\zeta,\cr
\delta \bar{c}^\alpha &=F^\alpha\zeta,\cr
\delta c^i &=-{1\over 2} \epsilon^{ijk}c^jc^k\zeta,\cr
\delta c^0 &=0.\cr
 }\eqno(2.2.8)$$
Since $F^\alpha$ is linear in the $U$ field (but not in the
$\pi$ field), a Ward identity similar to (2.1.13) can be derived
$$ k^2\Lambda_{L}^{W^+W^-}+ik^2M_0\sqrt{\xi_1^0\xi_2^0}
\tilde{\Lambda}^{W^+\pi^-}+{M_0^{2}\over 4}\xi_1^0\xi_2^0
\tilde{\Lambda}^{\pi^+\pi^-}=i\xi_1^0.\eqno(2.2.9)$$
(We have restored the two different gauge parameters $\xi_1^0$,
$\xi_2^0$.) The tilded Green functions in the previous
equation are defined as
$$ik_\mu\tilde{\Lambda}^{W^+\pi^-}(k^2)=
\int d^4x e^{-ikx}\langle 0\vert W_\mu^+(x)
[(v+\rho)U^-](0)\vert 0\rangle , \eqno(2.2.10)$$
$$\tilde{\Lambda}^{\pi^+\pi^-}(k^2)=
\int d^4x e^{-ikx}\langle 0\vert [(v+\rho)U^+](x)
[(v+\rho)U^-](0)\vert 0\rangle ,\eqno(2.2.11)$$
with $U^\pm={\rm Tr} \sigma^\pm U$. The Ward identity (2.2.11)
thus actually relates an infinite number of Green functions
when written in terms of the $\pi$ fields. These Ward
identities can be expanded in inverse powers of $v$ and
solved iteratively. In practice, however, it is more useful
to shift
to another set of gauge conditions which are linear in the
non-linear Goldstone fields $\pi$. Let us take instead
$$\eqalign{
F^{\pm} &= {1 \over \sqrt{\xi_{1}^W}} \partial^{\mu}
W_{\mu}^{\pm} -  M_{W} \sqrt{\xi_{2}^{W}} \pi^{\pm}, \cr
F^{3} &={1 \over \sqrt{\xi_{1}^3}} \partial^{\mu} W_{\mu}^{3}
-M_{W} \sqrt{\xi_{2}^{3}}\, \pi^{3}, \cr
F^{0} &={1 \over \sqrt{\xi_{1}^B}} \partial_{\mu} B^{\mu}
+(M_{Z}^2-M_{W}^2)^{1 \over 2} \sqrt{\xi_{2}^{B}}\,
\pi^{3}.}\eqno(2.2.12) $$
Then the Ward identities (2.1.13) and (2.1.14) remain strictly
valid with
$$i{k_\mu}\Lambda^{W^+\pi^-}(k^2)=\int d^4x e^{-ikx}\langle 0\vert
W^+_\mu(x) \pi^-(0)\vert 0\rangle ,\eqno(2.2.13)$$
$$\Lambda^{\pi^+\pi^-}(k^2)=\int d^4 x
e^{-ikx}\langle 0\vert \pi^+(x)
\pi^-(0)\vert 0\rangle .\eqno(2.2.14)$$
Even though in this gauge the expressions are formally the
same,
the bare Green functions themselves that appear in
these expressions differ numerically from those obtained in
the linear realization. The equality between both
realizations is only guaranteed at the level of $S$-matrix
elements[21], or for
connected
Green functions involving only gauge fields[19] (since they
have not changed in passing from  linear to
non-linear variables).

The lagrangian in this gauge, expanded
up to four fields, is given in Appendix D. Notice that
the Goldstone bosons have only derivative couplings and that
the coupling $\lambda$ now affects only the $\rho$ field. If the
mass of this radial excitation, which is to be identified with
$M_H$,
 is very large it makes sense to go one step further
and
 integrate $\rho$ out obtaining an effective lagrangian
that will reproduce the Standard Model at energies
much below $M_H$. This will be discussed in section 5 and 6.
For the time being let us return to the Standard
Model  in the linear realization.

\bigskip
\line{\bf 3. The Equivalence Theorem in the Standard Model\hfil}
 \medskip\noindent
We will start by  restricting ourselves to a physical process
that contains only one external longitudinal
vector boson and towards the end of the section we will
consider the generalization to an arbitrary number of vector
bosons.

Let us begin by recalling some elementary facts in the
Standard Model. The first point to remember is that
the gauge conditions (2.1.3) must be satisfied by the
{\it in} and {\it out} states. Therefore (and restricting ourselves
to the charged boson case)
$$\langle 0 \vert F^{\pm} \vert \psi \rangle = 0, \eqno(3.1)$$
$F^{\pm}$ being the gauge condition (2.1.4) and $\vert \psi\rangle$
some physical state. The second point to remember is that
the condition (3.1) does not determine the state completely.
We shall use a 2-vector notation to represent the
$W_\mu$ and $\omega$ contents of an external state. Thus
 a {\it in} state represented by
$$(\epsilon^\mu, 0) \qquad k \cdot \epsilon=0,
\eqno(3.2)$$
meaning that the asymptotic field representing
the Goldstone boson is zero,
and one represented by
$$
(\epsilon^\mu - {k^\mu \over M} \theta,{i \over  \sqrt{\xi_1\xi_2}}
\theta),\eqno(3.3)$$
where the asymptotic field representing the Goldstone boson is
multiplied by the second entry of the  vector (3.3), fulfill the
same gauge condition; they are the same physical state[22].
Since $k^2=M^2$, for massive vector
bosons it is not possible to take $\epsilon^\mu_L \propto
k^\mu$. Rather the polarization vectors associated to longitudinal $W$'s
are of the form
$$\epsilon^{\mu}_L={k^{\mu} \over M}+v^{\mu}.\eqno(3.4)$$
Therefore states that asymptotically correspond to
longitudinally polarized $W$'s , described by
polarization vectors $\epsilon^\mu_L$, cannot be completely
gauged away and traded for  Goldstone bosons. The
best one can do is to single out within the equivalence
class corresponding to a given physical state $\vert \psi\rangle$
two extreme cases. One is just (3.2), the other is obtained
by gauging away the $k^\mu$ part in (3.4), namely
$$(v^\mu, {i \over  {\sqrt{\xi_1\xi_2}}}).\eqno(3.5)$$
Finally, a third point to remember
is that at very high energies the $v^\mu$ part is unimportant.
Let us write explicitly the longitudinal polarization vector
for a $W$ particle moving forward in the $z$-axis
with momentum $p$ and energy $E$
$$\epsilon^\mu_L=({p\over M},0,0,{E\over M}).\eqno(3.6)$$
Therefore $\epsilon^\mu_L-k^\mu/M$ is of
${\cal O}(M/E)$, eventually negligible when compared
to the first term in (3.4) which is of ${\cal O}(E/M)$.
 Also, and for the same
reason, at high energies the scattering of longitudinal
components of vector bosons dominates over
transverse ones[2,20,23].

In a way eqs. (3.2) through (3.5) already encompass the Equivalence
Theorem. At high energies, the scattering of Goldstone
bosons has something to do with the scattering of longitudinally
polarized $W$'s. Obviously, this is only a qualitative
statement and one needs to go beyond this to make detailed
calculations. For instance, the Equivalence
Theorem appears to be
closely linked to the 't Hooft-Feynman gauge[24], in which
one sets $\xi_1=\xi_2=1$. What happens then for, say,
Landau gauge where Goldstone boson physics is more manifest? Can
(1.1)
be derived then? On the other hand, Green functions involving unphysical
states (such as those constructed with only
$\omega$ fields) need not be gauge invariant.
Only keeping the contribution from the $v^\mu$ part
ensures that and it is necessary to be well aware of this fact.

The answer to the first objection is of course that
eqs. (3.2)-(3.5) are formal relations valid for
bare quantities while we are interested in relations involving
$S$-matrix elements. Furthermore as the fields propagate
they mix among themselves since they have the same quantum
numbers. Let us
 start by deriving
a relation involving asymptotic fields. We shall
follow the approach of [9], but we will deviate at some point
in order not to introduce unphysical Green functions
involving ghosts. The condition (3.1) reads (for a negatively
charged {\it in} state $\vert \psi\rangle$)
$$ \eqalign{ 0&=\int d^4x e^{-ikx}\langle
0\vert \partial^\mu W_\mu^+(x)-M_0\sqrt{\xi_1^0\xi_2^0}
\omega^+(x)\vert \psi\rangle\cr
&=(ik^\mu,-M_0\sqrt{\xi_{1}^{0}\xi_{2}^{0}} )\int d^4x e^{-ikx}
\langle 0\vert (W^+_\mu(x),\omega^+(x))\vert
\psi\rangle^\top\vert_{k^{2}=M^{2}}. \cr}\eqno(3.7)$$
All quantities and fields appearing above
are understood to be bare ones.
Using now the reduction formula (see e.g. [25]) to relate Green
functions to amplitudes we get
$$K^M D_{MN}\langle
W^N\vert\psi\rangle^\top\vert_{k^{2}=M^{2}}=0,\eqno(3.8)$$ %
where $K^M= (ik^\mu,-M_0\sqrt{\xi_1^0\xi_2^0})$,
$$D_{MN}=\left(\matrix{ &D_{\mu\nu}^{W^+W^-}(k)
&D_\mu^{W^+\omega^-}(k)\cr
&D_\mu^{\omega^+W^-}(k) &D^{\omega^+\omega^-}(k)\cr}
\right),\eqno(3.9) $$
and
$$\langle W^N(k)\vert\psi\rangle=
\langle (W^{-\nu}(k), \omega^-(k)\vert\psi\rangle.\eqno(3.10)$$
(3.10) is not quite an amplitude because is not yet
contracted with $\epsilon_\nu$, but an amputated
Green function.
Everywhere we understand that the limit $k^2\to M^2$ has
to be taken.

A straightforward calculation that makes use of the
Ward identity (2.1.13) leads to
$$ i k^{\nu} \langle W^{-}_{\nu}(k) \vert \psi \rangle =
{k^2\over {M_0\sqrt{\xi_1^0\xi_2^0}}}
{{i\xi_1^0-k^2\Lambda_L^{W^{+}W^{-}}-M^2_0\xi_1^0\xi_2^0
\Lambda^{\omega^+\omega^-}}
\over{
{{i\xi_1^0+k^2\Lambda_L^{W^{+}W^{-}}+M^2_0\xi_1^0\xi_2^0
\Lambda^{\omega^+\omega^-}}}}}
 \langle \omega^-(k) \vert \psi \rangle. \eqno(3.11)$$
We have to set $k^2=M^2$ in the above expression. All quantities
and fields are still unrenormalized. We now proceed to write
everything in terms of renormalized quantities using the
renormalization constants described in Appendix A.
The external legs require special treatment.
In the on-shell scheme one commonly uses a minimal set of renormalization
constants that make finite all Green functions, but which do not
guarantee a unit residue for the vector boson propagators[16,26] (other
that the photon). To fix this problem one uses for the external legs
$\tilde{Z}_W$ and $\tilde{Z}_\omega$ defined in (A.4).

Introducing the renormalization
constants into (3.11)
 one finally gets a relation between
renormalized amputated Green functions
$$ i k^{\nu} \langle W^{-}_{\nu }(k) \vert \psi\rangle =
M {\cal C} \langle \omega^{-}(k) \vert \psi \rangle .
\eqno(3.12) $$
${\cal C}$  is given by
$$ {\cal C}=
\left( { \tilde{Z}_W \over {\tilde{Z}_\omega}}\right)
^{1/2} {k^{2} \over M^{2}}
{1 \over{
Z\sqrt{\xi_{1} \xi_{2}}}}
{i\xi_1- k^2 Z_{W}Z_{\xi_1}^{-1} \hat{\Lambda}_{L}^{W^{+}W^{-}}
 - \xi_{1} \xi_{2} M^2 Z Z_\omega Z_{\xi_1}^{-1}
\hat{\Lambda}^{\omega^{+}\omega^-}
\over
i\xi_1+ k^2 Z_{W} Z_{\xi_1}^{-1}\hat{\Lambda}_{L}^{W^{+}W^{-}}
 + \xi_{1} \xi_{2} M^2 Z Z_\omega Z_{\xi_1}^{-1}
\hat{\Lambda}^{\omega^{+} \omega^{-}}
}
\eqno(3.13) $$
As usual, the renormalized propagators appearing in (3.13) are to
be
evaluated at $k^2=M^2$. In the above
expression $Z= {\left( Z_{M} Z_{\xi_{1}} Z_{\xi_{2}} \right) }^{1/2} $.
The renormalization constants that show up in (3.13) appear
in combinations so as to make ${\cal C}$ finite. This
can be easily checked by recalling (2.1.15).
Notice that the renormalization of the external legs $
\tilde{Z}_W$ and $\tilde{Z}_{\omega}$ have been included in
${\cal C}$. The presence of this factor in the Equivalence Theorem
has been detected
before[8-10]. The expression given here is new, however. In [9,10]
${\cal C}$ is given in terms of Green functions involving ghosts.

The factor ${\cal C}$ as given above
is valid to
all orders. In the on-shell scheme and at the one
loop level all renormalization constants are expressable
in terms of bare self energies. If we particularize
to this order ${\cal C}$ becomes
$${\cal C}=\left(1 + {1 \over 2}(\delta
\tilde{Z}_W-\delta\tilde{Z}_\omega)+{1 \over 2 M^2}
 \left( \Sigma_T(M^2)
-\Sigma_L (M^{2}) - \Sigma_{\omega} (M^{2}) \right)
\right).
\eqno(3.14) $$
It is remarkable the
simplicity of this expression, which is valid in any gauge
(although the bare self energies themselves do depend on
the gauge). Furthermore the leading contribution
is just 1, reproducing
the naive arguments at the beginning of this
section using Feynman gauge.
Let's now analyze the properties of eq.(3.14).
Obviously ${\cal C}$
is finite in the Standard Model. It turns out to be independent of
$M_{H}$ too (see the $\Sigma$'s, ${\tilde Z}$'s in Appendix A).
At least at the
one loop level in the on-shell scheme, the
${\cal C}$ factor, and the Equivalence Theorem by extension, do
not have any sizeable corrections due to the scalar
sector and only contributions of
${\cal O}(g^2)$ appear.

As it stands, (3.12) relates finite, but yet unphysical,
quantities. We rather write it as
$$\epsilon^\mu\langle W_\mu^-(k)\vert \psi\rangle
= -{i } {\cal C}\langle \omega^-(k)\vert \psi\rangle
+v^\mu\langle W_\mu^-(k)\vert \psi\rangle,
\eqno(3.15)$$
which we denote as
$$\langle W_L^-\vert \psi\rangle=-i{\cal C}\langle\omega^-\vert
\psi\rangle+\langle \tilde{W}^-\vert \psi\rangle. \eqno(3.16)$$
The l.h.s. is now a physical $S$-matrix amplitude. The r.h.s. is
a sum of two pieces neither of which is physical. Both are
gauge dependent, but the gauge dependence cancels between
them and also thanks to the ${\cal C}$ factor (more on this
will be discussed later). All matrix elements
are to be computed at $k^2=M^2$. Since the Goldstone boson
mass in the Standard Model is a function of the gauge parameter
except in Feynman gauge the first matrix
element on the r.h.s. is also off-shell in general.

It is a simple matter to extend the above results to several
external longitudinally polarized $W$'s. We proceed iteratively,
applying the above procedure one by one to the external $W$'s
taking into account that the vector $K^M$ for an outgoing
$W^+$ or $W^-$ with momentum $k^{\mu}$ is $K^M=(ik^\mu,
-M\sqrt{\xi_1\xi_2})$,
while for an incoming $W^+$ or $W^-$ with momentum $k^{\mu}$
is $K^M=(-ik^\mu, -M\sqrt{\xi_1\xi_2})$.
We then end up with
$$\eqalign{
\langle W_L W_L W_L\dots \vert \psi\rangle= &(-i)^n {\cal C}^n
\langle \omega \omega \omega\dots \vert\psi\rangle\cr &+(-i)^{n-1}{\cal
C}^{n-1} \langle \tilde{W}\omega\omega\dots\vert \psi\rangle
+(-i)^{n-1}{\cal C}^{n-1}\langle\omega \tilde{W}\omega\dots\vert\psi
\rangle\dots \cr
&+{\cal O}((v^\mu)^2).\cr}\eqno(3.17)$$
For incoming $W_L$'s we have to replace the appropriate $-i$ by a $+i$.
For instance
$$\eqalign{A(W_L^+W_L^-\to W_L^+W_L^-)=& {\cal C}^4 A(\omega^+
\omega^-\to \omega^+\omega^-)\cr
&+i {\cal C}^3 A(\tilde{W}^+\omega^-\to \omega^+\omega^-)
+i {\cal C}^3 A(\omega^+\tilde{W}^-\to \omega^+ \omega^-)\cr
&-i {\cal C}^3 A(\omega^+\omega^-\to \tilde{W}^+\omega^-)
-i {\cal C}^3 A(\omega^+\omega^-\to \omega^+\tilde{W}^-) \cr
&+{\cal O}((v^\mu)^2).\cr}\eqno(3.18)$$
We shall lump together the four pieces with one $W_\mu$ external
field contracted with $v^\mu$ and three $\omega$'s under
the symbol $A(\tilde{W}\omega\omega\omega)$
$$ \eqalign{A({\tilde W} w w w)=&
i {\cal C}^3 A(\tilde{W}^+\omega^-\to \omega^+\omega^-)
+i {\cal C}^3 A(\omega^+\tilde{W}^-\to \omega^+ \omega^-)\cr
&-i {\cal C}^3 A(\omega^+\omega^-\to \tilde{W}^+\omega^-)
-i {\cal C}^3 A(\omega^+\omega^-\to \omega^+\tilde{W}^-).}\eqno(3.19)$$
Of course, if the
Equivalence Theorem is ever going to be useful we must be able to stop
the expansion in (3.18) at some, preferably early, point. How far we
need to go
depends both the energy of the process (since further contributions are
suppressed by additional powers of the energy) and on the precision required
(are we merely interested in the limit where $\lambda$ is large, or
${\cal O}(g^2)$ corrections need to be taken into account?).

We just saw that $v^\mu$ is suppressed by a factor $M^2/E^2$ with
respect to $\epsilon_L^\mu$. The additional terms,
$A(\tilde{W}\omega\omega\omega)$ are therefore suppressed by this same
factor with respect to $A(\omega\omega\omega\omega)$. In a renormalizable
theory the latter behaves for large $E$ as a constant, at worst. Thus
the additional term should indeed be of ${\cal O}(M^2/E^2)$. However,
this need not be the case in a non-renormalizable effective theory. It is
quite admissible that in an effective theory the amplitude
grows as $E^2/v^2$, $E^4/v^4$, etc. In this case the additional
terms in $A(\tilde{W}\omega\omega\omega)$ do not vanish when $M^2/E^2
\to 0$. The non-unitary growth with the energy cannot continue
indefinitely, of course. At some point the expansion in powers
of $E^2/v^2$ simply breaks down and beyond that point the additional
terms $A(\tilde{W}\omega\omega\omega)$ will eventually tend to zero
as $M^2/E^2\to 0$.  The large $M_H$ limit in the Standard Model
amplitudes is a good example to illustrate the previous point. If
$M_H$ is very large, for $E\ll M_H$ the tree level amplitudes
grow (for a while) as $E^2/v^2$ and the additional terms are
non-negligible even for reasonably large energies.

\bigskip
\line{\bf 4. Applications
of the Equivalence Theorem in the Standard Model.\hfil}
\medskip\noindent
In this section we will explicitly check the validity of the
Equivalence Theorem at tree level in the Standard Model
including the first subleading
corrections in (3.18), which
have not been computed before. Our results clarify some
misunderstandings.
As we just discussed,
although
it is frequently stated that the next to leading terms
$A(\tilde{W}\omega\omega\omega)$ are of ${\cal O}(M/E)$
[2,11-12]
this is not always true. In particular for the scattering $W_{L} W_L
\to  W_L W_{L}$ when $M_{H} > s $ they give
sizeable contributions
at small angles. (In [4] the need to consider the subleading
amplitudes is pointed out, but the analysis presented there
only scratches the surface of the problem.)
Another point we would like to stress is
that the Equivalence Theorem,  as stated in
(3.17), is an exact result with no need of
taking any limit whatsoever
[11,24]. Finally, with regards to some
difficulties
with Lorentz invariance and the fact that one is dealing with
longitudinally polarized particles which have been recently
reported in [12], they do not show up if one always works in
a kinematical region where {\it all} energies are much greater than $M$.
All calculations have been carried out in t'Hooft-Feynman gauge.
The processes $A(W^{+} W^{-} \rightarrow W^{+} W^{-})$ and $A(W^{+}
W^{+}
\rightarrow W^{+} W^{+})$ have been calculated previously in [27] and
[28], respectively.

\bigskip
\line{\bf 4.1. $ W^{+} W^{-} \rightarrow W^{+} W^{-} $\hfil}
\medskip\noindent
The exact amplitude for the scattering of $W_{L}$ is given by the set
of diagrams in Fig. 1
$$\eqalign{A( W_{L}^{+} W_{L}^{-} \rightarrow W_{L}^{+} W_{L}^{-} )= &
{g^2 \over 4 M^{4} x^2} \biggl[ (4 M^{2} + x) (4 M^{2} (t^2 +
t x) + t^2 x + 4 t x^2 + x^3 ) \cr
&+M^{2} {{(2 M^{2} (2 t + x) + tx )}^{2} \over M_{H}^{2} - t} +
M^2 x^{2} {{(2 M^{2} + x)}^{2} \over M_{H}^{2} -s } \cr
&+ \left( {s_{w}^{2} \over t} + {c_{w}^{2} \over t - M_{Z}^{2}} \right)
\biggl( 4 M^{4} (x - 2t)
(2 t^2 - tx -2 x^2)  \cr  &  - 16 M^{6}{ (2t + x)}^{2}
 - 8M^{2} (t^{3} x - t x^{3}) - t^{2} x^{2} (t + 2 x)\biggr) \cr
& \left. - \left( {s_{w}^{2} \over s} + {c_{w}^{2} \over s - M_{Z}^{2}}
\right) x^{2} ( 2 t + x){(x + 6 M^{2} )}^{2} \right].
}\eqno(4.1.1)$$
$s,t,u$ are the Mandelstam variables defined by
$s=(p_{1}+p_{2})^{2}$, $t=(p_{1}-k_{1})^{2}$ and
$u=(p_{1}-k_{2})^{2}$. We have introduced the variable $x=s-4M^2$ to
simplify the expressions. In the CM frame
$$\eqalign{ &p_{1}=( E,0,0,p ) \cr &p_{2}=( E,0,0,-p ) }\qquad
\qquad
\eqalign{&k_{1}=( E,p \sin\theta,0,p \cos\theta ) \cr
&k_{2}=( E,-p \sin\theta ,0,-p \cos\theta ) }\eqno(4.1.2)$$
and
$$\eqalign{ &\epsilon_{L}(p_{1})={1 \over M} ( p,0,0,E ) \cr
&\epsilon_{L}(p_{2})={1 \over M} ( p,0,0,-E ) } \qquad \qquad
\eqalign{&\epsilon_{L}(k_{1})={1 \over M} ( p,E \sin \theta,0,E
\cos \theta ) \cr
&\epsilon_{L}(k_{2})={1 \over M} ( p,-E \sin \theta,0,-E \cos \theta
) }\eqno(4.1.3)$$
The following kinematical relations hold:
$p^{2}={x/4}$ and $\cos \theta = 1 + 2t/x$.

The Goldstone boson amplitude is given by the
diagrams of Fig. 2. The calculation is simpler (this is one of the
main assets of the Equivalence Theorem, of course). We set
${\cal C}=1$ for the time being
$$\eqalign{
A(\omega^{+} \omega^{-} \rightarrow \omega^{+} \omega^{-})
=&{g^2} \left[ {M_{H}^{2} \over 8 M^{2}} {(M_{H}^{2} (s+t)
-2 s t) \over (M_{H}^{2} - s)(M_{H}^{2} - t)} \right. \cr
&\left. + (4 M^{2} - 2 t -
s)
 \left( {(c_{w}^{2}-s_{w}^{2})^{2} \over 4 c_{w}^{2} } {1 \over s -
M_{Z}^{2} } + {s_{w}^{2} \over s} \right) \right]+ \left( s
\leftrightarrow t \right) } \eqno(4.1.4) $$
The first correction to the `naive' Equivalence Theorem corresponds
to the diagrams in Fig. 3. Their contribution to the
r.h.s. of (3.18) is
$$\eqalign{
A({\tilde W} \omega \omega \omega)=
&g^2 \left[ {M_{H}^{2} \over  M^{2}} \left( {x (M^{2} - t)
+t \sqrt{xs}  \over x (M_{H}^2 -t)} + {(M^{2}-s) + \sqrt{xs} \over
M_{H}^2 -s} \right) \right. \cr
&\left. + 4 s_{w}^{2}\left( {( x^2 + x (s+t) - \sqrt{xs} (2 x + t)
 ) \over 2 x t}
+{ ( x^2 + 2 t x - \sqrt{xs} (x + 2 t )
 \over 2 x s}  \right ) \right] \cr
&+g^{\prime 2} {c_{w}^2 - s_{w}^{2} \over  x} \left[
{(x + 2 t) (\sqrt{xs} -x ) \over s-M_{Z}^{2} }
+ { (2 x + t) \sqrt{xs} - x (s+t+x) \over t- M_{Z}^{2}} \right]}
\eqno(4.1.5) $$
In order to check analytically the validity of the
Equivalence Theorem let us expand (4.1.1) to (4.1.5) in inverse powers
of the energy. Except for $\cos \theta\sim 1$, $s\to\infty \Rightarrow
-t\to \infty$. Therefore we perform a double expansion in
$M^2/s$ and $M^2/t$
$$\eqalign{A(
 W_{L}^{+} W_{L}^{-} \rightarrow W_{L}^{+} W_{L}^{-} )=
& -{g^2 \over
4}{ M_{H}^{2} \over M^{2}} \left[ {t \over t-M_{H}^{2}}
+ {s \over
s-M_{H}^{2}} \right]
 - {g^2 \over 2 c_{w}^2} {s^2 +t^2+ st \over st} \cr
&+g^2 {M_{H}^{2} \over s} {2 M_{H}^2 t - s(s+t) \over (M_{H}^2 -s)
(M_{H}^{2} - t) }
+{\cal O}({M^{2}/s}, M^2/t ),  }\eqno(4.1.6)$$
$$\eqalign{A(\omega^{+} \omega^{-} \rightarrow \omega^{+} \omega^{-})=&
 -{g^2 \over
4}{ M_{H}^{2} \over M^{2}} \left[ {t \over t-M_{H}^{2}}
+ {s \over
s-M_{H}^{2}} \right]
- {g^2 \over 2 c_{w}^2} {s^2 +t^2+ st \over st} \cr
&+{\cal O}({M^{2}/ s},M^2/t),  }\eqno(4.1.7) $$
and, finally, (4.1.5) becomes
$$
A({\tilde W} \omega \omega \omega)=
{g^2 }{M_{H}^{2} \over s} {2 M_{H}^2 t - s(s+t) \over (M_{H}^2
-s) (M_{H}^{2} - t)}
+{\cal O}({M^{2}/s},M^2/t).\eqno(4.1.8) $$
We have kept the
complete Higgs structure in the denominator. Adding (4.1.8) and (4.1.7)
reproduces (4.1.6). Notice that in the large $M_H$ limit the additional
correction to the `naive' Equivalence Theorem is of ${\cal O}(1)$ in
the $1/E$ expansion  and this in spite of
the explicit $v^\mu$ suppression factor. Indeed, although this
part of the
amplitude is suppressed
with respect to the leading term (4.1.6) by one power of $M^2/E^2$,
the amplitudes grow as $E^2/v^2$ in the large Higgs mass
limit --- a hint of the perturbative problems with unitarity in
the Standard Model.
The additional piece $A(\tilde{W}\omega\omega\omega)$
is ${\cal O}(g^2)$ and
thus definitely subleading w.r.t.
(4.1.7)
in accordance with our expectations,
but not negligible in any case.

In the opposite extreme,
if the Higgs is light, (4.1.5) is
${\cal O}(g^2M_H^2/E^2)$, again subleading with respect to
(4.1.7) , which is ${\cal O}(\lambda)$. But now
$A(\tilde{W}\omega\omega\omega)$ does indeed vanish as $E\to\infty$,
in accordance with the `naive' statement of the Equivalence Theorem
(1.1). In either case
we have not included the factor ${\cal C}$. This would be
required if we desire to work with a ${\cal O}(\lambda g^2)$
or ${\cal O}(g^2 E^2/v^2)$ accuracy, but that would also require
computing loop corrections.

\bigskip
\line{\bf 4.2. $ W^{+} W^{+} \rightarrow W^{+} W^{+} $\hfil}
\medskip\noindent
The diagrams entering this
amplitude are similar to those of Fig. 1,  but exchanging the
$s$ and $u$ channels. The tree level results are
$$\eqalign{A(W_{L}^{+} W_{L}^{+} \rightarrow W_{L}^{+} W_{L}^{+} )=
&{g^2 \over 4 M^{4} {(t+u)}^{2}} \biggl[
 16 M^{4} t u
-2 M^{2}(t+u)(t^2 + 6tu + u^2)  \cr
&+ {1 \over 2} (t+u)^2 (t^2 + 4tu +u^2) -
M^{2}{(2 M^2 (t-u) + tu +
u^2)^2 \over u - M_H^2 } \cr
&+ \left( {c_w^2 \over u- M_Z^2 }
+ {s_w^2 \over u} \right)
\left( 16 M^6 (t-u)^2
+ 8 M^2 tu (t+u)(t+ 2u)
 \right. \cr
& \left.
- 4 M^4 (t+3 u) (2 t^2 +3 tu - u^2 )
 -u^2 (t+u)^2(2t +u) \right) \biggr] \cr
&+(t
\leftrightarrow u  ), }\eqno(4.2.1) $$
$$\eqalign{A(\omega^+ \omega^{+} \rightarrow \omega^{+}
\omega^{+})=&{g^2} \left[ {M_{H}^{2} \over 8 M^{2}} {(M_{H}^{2} (t+u)
-2 t u) \over (M_{H}^{2} - t)(M_{H}^{2} - u)} \right. \cr
&\left. + (4 M^{2} - 2 t -
u)
 \left( {(c_{w}^{2}-s_{w}^{2})^{2} \over 4 c_{w}^{2} } {1 \over u -
M_{Z}^{2} } + {s_{w}^{2} \over u} \right) \right]+ ( t
\leftrightarrow u ),}  \eqno(4.2.2) $$
and, for the leading correction to the `naive'  Equivalence Theorem,
$$\eqalign{
A({\tilde W} \omega \omega \omega)=&
{g^2} \biggl[ {M_{H}^{2} \over  M^{2}}{\left(
M^{2} (t+u) - tu -u^2 -u \sqrt{sx} \right)
\over (M_{H}^{2} - u) (t+u)}
+4 \biggl( {s_{w}^{2} \over 2 u (t+u) }\cr
&+ {s_{w}^2 (c_{w}^{2} -
s_{w}^{2} )\over 4 c_{w}^2 (M_{Z}^{2}-u) (t+u) } \biggr)
\biggl( -4 M^{2} (t+u)
+ 2 t^{2} + u^{2} + 3 t u \cr
&+ (2 t +u) \sqrt{sx} \biggr)
\biggr] + \left( t \leftrightarrow u \right).} \eqno(4.2.3)
$$
As in the previous case, in order to ease the comparison between
these amplitudes we shall
expand them up to ${\cal O}(M^2/E^2)$. For simplicity we work away
from the forward and backward regions. Then both $t$ and $u$ are
large and
$$\eqalign{A(
 W_{L}^{+} W_{L}^{+} \rightarrow W_{L}^{+} W_{L}^{+} )=
& -{g^2 \over
4}{ M_H^2 \over M^2} \left[ {t \over t-M_H^2}
+ {u \over
u-M_H^2} \right]
 - {g^2 \over 2 c_w^2} {t^2 +u^2+ tu \over tu} \cr
&-g^2 {M_{H}^{2} \over (t+u)} {(t-u)^{2} \over (M_{H}^2 -t)
(M_{H}^{2} - u) }
+{\cal O}(M^{2}/t,M^2/u),  }\eqno(4.2.4)
 $$
$$\eqalign{A(\omega^{+} \omega^{+} \rightarrow \omega^{+} \omega^{+})=&
 -{g^2 \over
4}{ M_{H}^{2} \over M^{2}} \left[ {t \over t-M_{H}^{2}}
+ {u \over
u-M_H^2} \right]
- {g^2 \over 2 c_{w}^2} {t^2 +u^2+ tu \over tu} \cr
&+{\cal O}(M^2/t,M^2/u),}\eqno(4.2.5) $$
and for the additional piece in this limit we get

$$
A({\tilde W} \omega \omega \omega)=
-g^2 {M_H^2 \over (t+u)} {(t-u)^2 \over (M_H^2 -t)
(M_H^2 - u) }
+{\cal O}(M^2/t,M^2/u).  \eqno(4.2.6) $$
The addition of eq.(4.2.5) and (4.2.6)
reproduces eq. (4.2.4). Now, however, (4.2.6) is ${\cal
O}(g^2 E^2/M_H^2)$  for large values of the Higgs mass.
Since the leading
term is still ${\cal O}(E^2/v^2)$,
the leading corrections to the
`naive' Equivalence Theorem are down by a factor
$M^2/M_H^2$ which is actually smaller than
$M^2 / E^2$ in this limit. For a light Higgs, the
correction is
${\cal O}(g^2 M_H^2/E^2)$, to be compared with the
leading  ${\cal O}(\lambda)$ contribution, this time
in accordance with the usual counting.

Of course,
both in this case and in the previous one, the additional contributions
can be greatly enhanced by some kinematical reasons, e.g.
in the vicinity of the Higgs pole. We discuss this issue
in more detail in the next subsection.

\bigskip
\line{\bf 4.3. Domain of Applicability\hfil}
\medskip\noindent
Let us now analyze more carefully the improvement on
the Equivalence Theorem that is brought about by keeping the
additional terms in (3.18), such as (4.1.5) and (4.2.3).

We will discuss here the scattering
$W_{L}^{+} W_{L}^{-} \rightarrow W_{L}^{+} W_{L}^{-}$ whose
tree level results have been described in
section 4.1.  We have plotted this amplitude for three different angles
($\theta=\pi / 16, \pi / 4 , 3 \pi / 4$) (Fig. 4). We take
as physical input
$M_{Z}$, $M$, $M_{H}$ and $\alpha$ and work in the region
$2 M \ll E \ll M_{H}$.
We have taken  $M_{H}=1$ TeV.

The solid line corresponds
to the exact tree level result for $W_L$ scattering.
The short-dashed line corresponds to the $g=g^\prime=0$ limit, which is
the standard approximation in the literature (see [2,28] and the second
reference in [27]). Notice that the corrections to the `naive' Equivalence
 Theorem are
always proportional to $g^2$. From Fig.4 it is clear that setting $g=0$
is a very crude approximation,
particularly at small angles, and that the difference does
not go to zero as $E\to\infty$ since
the additional
terms that correct the `naive' Equivalence Theorem are not of ${\cal
O}(M^2/E^2)$
but rather {\it down} by a factor $M^2/E^2$ with
respect to the leading contribution, which is
quite different. In fact their effect may be quite
sizeable.

The corrections for $g\neq 0$ have two origins. On the one hand
the first term on the r.h.s. of (3.18), which corresponds
to the `naive' Equivalence Theorem, gets contributions from the
exchange of $\gamma$ and $Z$. Adding these corrections
(dashed-dotted line)
improves the agreement with the exact result substantially
but still fails to reproduce the scattering amplitude of longitudinal
$W$'s in many kinematical regions. When we finally add the correction
contained in (4.1.5) the result (represented by a long dashed line) is
practically indistinguishable from the exact one for all kinematical
regions. In fact it is so close as to become invisible for most
angles. The terms proportional to $(v^\mu)^2$ and beyond in (3.18) are
obviously unimportant.

At this stage one
should stress that it is totally unnecessary to expand the
amplitudes in inverse powers of $M_H$ to verify the consistency
of the Equivalence Theorem as is sometimes done [24]. The
full analytical structure of the Higgs propagator is well reproduced
(including ${\cal O}(M^2/E^2)$ corrections) by adding the
terms proportional to $v^\mu$.
It is also illustrative to consider a complete $1/E$ expansion of the
$Z$ propagators contained in the amplitudes. The results (represented
by a dotted line in the Fig. 4. (b)) are obviously much worse.

\bigskip
\line{\bf 5. Effective Chiral Lagrangian\hfil}
\medskip\noindent
So far we have applied the Equivalence Theorem at tree level
in the minimal Standard Model. It is far more interesting
to go beyond this level and apply the above results at higher
orders in the perturbative expansion or, better, to use
the connection that it provides between scattering of longitudinal
$W$'s and Goldstone bosons to set bounds on new physics in the longitudinal
sector. The level of precision required in the latter case is
typically also that of a radiative correction since tree level type
modifications are by now excluded.

It is convenient and economical to treat the minimal Standard Model and
other theoretical possibilities on the same footing[29]. Provided
that the Higgs mass is sufficiently large, this
can be achieved by working with an effective chiral lagrangian.
This consists of a collection of operators with the required
symmetry properties of $SU(2)_L\times U(1)_Y$ local gauge invariance
and containing the Goldstone bosons of the $SU(2)_L\times
SU(2)_R\to SU(2)_V$ global symmetry breaking. These two conditions
greatly restrict the possible operators. Chiral and gauge invariance
force the interactions to be derivative and effective operators
can be classified according to the powers of momenta.

The most general Electroweak  Chiral Lagrangian
up to ${\cal O}(p^4)$ is of
the form
$$
{\cal L}^{eff}=-{1 \over 2} {\rm Tr} W_{\mu\nu} W^{\mu\nu}
-{1 \over 4} B_{\mu\nu} B^{\mu\nu}+{v^2 \over 4} {\rm Tr} D_{\mu} U^{\dag}
D^{\mu} U + \sum_{i=0,13} a_i {\cal L}_{i}
+{\cal L}_{GF}+{\cal L}_{FP}
\eqno (5.1)
$$
The complete list of operators is given in Appendix C. The $SU(2)_L\times
U(1)_Y$ gauge symmetry is realized non-linearly at the level of the
Goldstone bosons (see (2.2.3) and (2.2.4)).

For a sufficiently large Higgs mass, the minimal Standard Model is just
a particular case of (5.1). The actual value of the set of
coefficients $\{a_i\}$ that correspond to the minimal Standard
Model is obtained from (2.2.5), (2.2.6) and (2.2.7) after
integrating out the $\rho$ field. The safest way to obtain
their value
is through the matching conditions[7,18-19,30],
in which one requires equivalent descriptions in terms
of fundamental and effective theories, therefore determining the
values of $\{a_i\}$. The matching is done in perturbation theory
at the one loop level.
Although it would be clearly desirable to
go beyond this, no results are available at present.

The matching of both theories
requires some care[19] due to the subtleties of gauge
invariance and gauge-fixing.
The matching could {\it a priori} be carried out at different
levels: $S$-matrix elements, connected Green functions,
effective action, etc.
The softest requirement is to demand equal physical
$S$-matrix elements. This method is bound to work in
all cases[21], but,
given the large number of possible operators in the effective
theory, it is
cumbersome.
It is important to remember that
whenever one takes advantage of  an
effective theory to describe a physical  system one is
using a different (sometimes coarser) set
of variables to describe the Hilbert space of the system.
There is absolutely no guarantee that anything other
than observables should agree when using two different
sets of variables.
In fact, it may not even be possible to pose the question
meaningfully. Therefore, the use of Green functions to
perform the matching is, generally speaking, ruled out.
Fortunately, for the case at hand
it was shown in [19] that
it is possible (and actually simplest) to match the
fundamental and effective theories for renormalized connected Green
functions containing only gauge fields. This is because
the gauge fields are insensitive to the way the
scalar sector is parametrized.
On the other hand, requiring the matching at the
level of the effective action (generating functional
of 1PI diagrams) as it is sometimes done is not consistent.

When matching renormalized connected Green functions
between the fundamental and the effective theory one must be careful
to project out the longitudinal parts. These
are gauge dependent and there is
no guarantee that
the gauge in the fundamental and in the effective
theory are the same, since by definition this is not observable.
Thus it is not guaranteed that with the set of gauge invariant
operators
contained in (5.1) one should be able to reproduce the
longitudinal parts of the Green functions. Rather one
will in general need to consider also other BRS invariant
(but not gauge invariant) operators of the right dimensionality
to proceed with the matching.
In practice this means
that there are some coefficients in the lagrangian (5.1) which
cannot be determined.

In the Effective Chiral Lagrangian there are operators
that either vanish or simply reduce to other operators
when the equations of motion are used. They correspond to the
coefficients
$a_{11}$, $a_{12}$
and $a_{13}$. Therefore it is quite clear that working at
the one loop level they will never be determined via
$S$-matrix elements. It turns out that they cannot be determined
via renormalized Green functions either because they contribute to
the gauge longitudinal parts which, as discussed, require
additional BRS invariant operators to match and we end up
with more unknowns than matching equations. ($a_{11}$, ...
contribute to other Green functions as well, but these
involve Goldstone bosons, which are also unphysical)
Then, the
longitudinal part of Green functions cannot be unambiguosly fixed
in an effective theory. Yet the Equivalence Theorem is
basically concerned with longitudinal parts. Is the Equivalence
Theorem in jeopardy in an Effective Chiral Lagrangian?
We will return to this crucial point in the next section.

If one is interested in reproducing the minimal Standard Model
at tree level for energies $E\ll M_H$
it is enough to keep $a_5$
and set
$$ a_5^{tree}={v^2\over 8 M_H^2}.\eqno(5.2)$$
At the one loop level one requires the full expression
for $a_5$ and the other coefficients, which can be found in
[7,18-19,30]. The natural expansion parameter in an
Effective Chiral Lagrangian being $E^2/v^2$ (or rather
$E^2/{(4\pi v)}^2$), the effective theory
lends itself very easily
to the sort of energy expansion that is a characteristic
of the Equivalence Theorem.

The amplitude for the scattering of longitudinally polarized
$W$'s
takes the symbolic form
$$\eqalign{A=& (b^{(0)}_1 g^2 + b^{(0)}_2 g^2 {M^2\over E^2}
+\ldots)(1+{\cal O}({g^2\over {16\pi^2}}))\cr
& +{E^2\over v^2}(b^{(2)}_1 + b^{(2)}_2{g^2\over 16\pi^2}+\ldots)\cr
& +{E^4\over {16\pi^2 v^4}}(b^{(4)}_1
+b^{(4)}_2{g^2\over 16\pi^2}+\ldots )\cr
& +\ldots} \eqno(5.3)$$
The first line on the r.h.s. of (5.3) has its origin
in tree level exchange of vector bosons, once expanded
in powers of $E$.
The interesting physics is in the $a_i$ coefficients that
are contained in the constants $b^{(4)}_1$ and $b^{(2)}_2$. Clearly,
to
make any definite statements on these coefficients via the Equivalence
Theorem we need to be able to compute the r.h.s. of (3.18)
(or rather its counterpart in an Effective Chiral Lagrangian)
with enough accuracy. In previous sections we have seen
that the `naive' Equivalence Theorem has corrections
that modify the leading term by factors of ${\cal O}(M^2/E^2)$.
If this also holds in an effective theory, we need
to assume that $M^2/E^2$ is small, for the Equivalence Theorem
to be of practical use. On the other hand, it must be
satisfied that $E^2\ll 16\pi^2 v^2= 64 \pi^{2} M^2/g^2$. This
seems to provide a reasonably large window of applicability. Of course
this window must get bigger when we include more and more
corrections on the r.h.s. of (3.18). It is our
contention that adding the first non-leading corrections is
enough for practical applications.

If we keep in our effective lagrangian terms of ${\cal O}(p^4)$
at most and work at the one loop order only terms  of up to ${\cal
O}(E^4/v^4)$
will be generated in the different amplitudes appearing in (3.18). Since
there is a suppression factor $M^2/E^2$
due to the $v^\mu$ factor, the correction to the `naive' Equivalence
Theorem will produce terms of ${\cal O}(g^2)$ and ${\cal
O}(g^2E^2/v^2)$
as well as terms that are suppressed by powers of $M^2/E^2$. Further
corrections (terms with two $v^\mu$ or more) would produce contributions
either of
${\cal O}(g^4)$ or right away suppressed
by powers of $M^2/E^2$. Clearly at large energies (but still much less
than $4\pi v$), the relevant contributions will be contained in the
two terms that we keep on the r.h.s. of (3.18).
Since factors of $4\pi$, etc may be relevant, our claim  can
only be fully justified by a detailed calculation which is presented
in section 7.

\bigskip
\line{\bf 6. The Equivalence
Theorem and the Effective Chiral Lagrangian\hfil}
\medskip\noindent
In the effective theory we shall use the
same gauge condition as in the non-linear realization of the
minimal Standard Model
$$
F^{\pm} = {1 \over \sqrt{\xi_1}} \partial^{\mu}
W_{\mu}^{\pm} -  M \sqrt{\xi_2} \pi^{\pm}.
\eqno (6.1)
$$
Since the derivation of the Equivalence Theorem hinges on
the use of Ward identities, it is not difficult to see
that all steps hold
the case of an Electroweak Chiral Lagrangian. Therefore
$$\eqalign{A(W_L^{+} W_L^{-} \to W_L^{+} W_L^{-})= & {\cal C}^4
A(\pi^{+} \pi^{-}\to \pi^{+}\pi^{-})\cr
&+i{\cal C}^3 A(\tilde{W}^{+}\pi^{-}\to \pi^{+}\pi^{-})
+i {\cal C}^3 A(\pi^{+}\tilde{W}^{-}\to \pi^{+} \pi^{-})\cr
&-i{\cal C}^3 A(\pi^{+}\pi^{-}\to \tilde{W}^+\pi^{-})
-i{\cal C}^3 A(\pi^{+}\pi^{-}\to \pi^{+}\tilde{W}^-) \cr
&+{\cal O}((v^\mu)^2).\cr}\eqno(6.2)$$
In addition, in
obtaining the formal expression for the ${\cal C}$ factor
nothing depends on the
particular theory we are using; it is just a consequence of the Ward
identities of the theory. Being completely
general, eq. (3.14) carries over to the Effective Chiral Lagrangian
merely replacing $\Sigma_\omega$ and $Z_\omega$ by
$\Sigma_\pi$ and $Z_\pi$. Because ${\cal C}$ is finite when $M_{H}
\to \infty$ in the minimal Standard Model, it stays finite in an
Electroweak Chiral Lagrangian[6].

It is perhaps useful to start our discussion
by choosing the values of the $a_i$ coefficients [7,18-19,30]
that reproduce the minimal Standard Model. Let us emphasize that
`reproducing the Standard Model'
does not imply that bare self-energies and renormalization
constants have to be numerically equal to those used in section 4. In
general they will not be. But physical amplitudes will.
The bare self-energies will have two types of contributions:
from the ${\cal O}(p^2)$ lagrangian (including in this the gauge part),
entering both at tree level and at the one loop level, and from the
${\cal O}(p^4)$ lagrangian,  entering only at tree level, according
to the usual chiral counting rules.
The ${\cal O}(p^4)$  contribution
to the self-energies is given in Appendix C.

It is quite instructive to repeat the verification
of the Equivalence Theorem at tree level in the minimal
Standard Model in the language of the Effective Chiral
Lagrangian. The amplitudes $A(W^+ W^-\to W^+W^-)$
or $A(W^+W^+\to W^+W^+)$ come out exactly as in (4.1.1)
and (4.2.1), except that they appear expanded in inverse powers of
$M_H$. The left hand side of (3.18) changes completely.
There is a reshuffling of different contributions between
the leading and the subleading terms. In particular, one
can easily see that the amplitudes $A(\pi^+\pi^-\to
\pi^+\pi^-)$ and $A(\pi^+\pi^+\to
\pi^+\pi^+)$ have
changed. This should be no surprise as  they
are not physical amplitudes and may perfectly be
different in the Standard Model and in its Effective Chiral
Lagrangian (compare formulae  (4.1.7) and (7.1.7)).

At the one loop level we have to use the values derived in
[7,18-19,30]
for the $\{ a_i\}$, or simply keep them arbitrary if we wish to
parametrize
different alternatives to the minimal Standard Model. At this order
we will face the problem of the uncertainties in the longitudinal
components of the Green functions we have alluded to before.
To be definite we will pick a particular operator that
contributes to the longitudinal parts, such as the one with coefficient
$a_{11}$, and follow its track through the different contributions
in (6.2). $a_{11}$ might, on dimensional grounds, appear in principle as a
contribution
of ${\cal O}(E^4/v^4)$ through the diagram with four $\pi$'s. However,
the structure of the operator ${\cal L}_{11}$ is such that the
contribution is ${\cal O}(g^2E^2/v^2)$.
In addition $a_{11}$ may show up as a contribution
via radiative corrections to any of the two amplitudes on the
r.h.s. of (6.2) or via ${\cal C}$. The contribution would be
in either case of ${\cal O}(g^2E^2/v^2)$. In conclusion, although
$a_{11}$ appears almost
everywhere in the course of the calculation, at the end of the day
$a_{11}$ should drop from the r.h.s. of the Equivalence Theorem
because $a_{11}$ cannot appear on the l.h.s. given that $A(W_LW_L\to
W_LW_L)$ is
physical and we are working at the one loop level ($a_{11}$ could
conceiveably appear at the two loop order).
 Let us see this in detail.

The diagrams to compute are depicted in Fig.5. We shall work in Landau
gauge, but we have checked the cancellation of the gauge dependence.
The pion amplitude has two types of contributions proportional to
$a_{11}$. On the one hand, the diagram (a) of Fig. 5 gives
$$ - {4  g^2 \over v^2}(s+t) a_{11} .\eqno(6.3) $$
(Only the part of the amplitude proportional to $a_{11}$ is presented
here.) On the other hand there is a contribution to the
external legs represented by (b). There are four such diagrams.
Adding the
four of them  one gets
$$
 {4  g^2 \over v^2}(s+t) a_{11} .\eqno(6.4) $$
Diagram (c) vanishes in Landau gauge. The total contribution
from diagrams with external Goldstone bosons vanishes.
Finally, the amplitude with one $W^{\pm}$ and three Goldstone
bosons gets contributions from diagrams
(d) and (e), which respectively give
$$
-{4  g^2 \over v^2}(s+t) a_{11} \qquad {4  g^2 \over v^2}(s+t) a_{11}
.\eqno(6.5) $$ %
The bare amplitudes do not depend on $a_{11}$. The renormalization
constants and self-energies (Appendices A and C) entering ${\cal C}$ do
depend on $a_{11}$, however. Therefore both ${\cal C}$ and the
renormalized amplitudes are potentially dependent on $a_{11}$. Yet, in
Landau gauge, which we are using, ${\cal
C}$ turns out to be $a_{11}$-free and so are the renormalized amplitudes.
The moral is that the Green functions that appear
in the formulation of the Equivalence Theorem are one by one potentially
ambiguous but the ambiguities drop in physical quantities. In practice
there is no need to go through the painstaking process of constructing
BRS invariant operators, matching them and keeping track of these
spureous longitudinal parts.

\bigskip
\line{\bf 7. Applying the Equivalence Theorem to the
Effective Chiral Lagrangian \hfil}
\medskip\noindent
As discussed in section 5, one is working here
within an energy expansion. On the other hand, the Equivalence
Theorem
implies an  expansion in inverse powers of the energy.
It is obvious that
these two expansions can give at best a window of applicability.  The
question whether this window is of zero or negligible width  has
been recently raised in [4,13]. These authors have considered
the $g=0$
approximation. Numerical analysis[13] show that then the
Equivalence Theorem holds only for
very high energies, sometimes
higher than the regions where chiral perturbation theory can be
trusted.
We would like now to substantiate the claim
that
keeping the first leading corrections to the `naive' Equivalence Theorem
is enough to restore the agreement.

\bigskip
\line{\bf 7.1. $A( W^{+} W^{-} \rightarrow W^{+} W^{-} )$ \hfil}
\medskip\noindent
Here we will be interested in analyzing this amplitude, already studied
in the minimal Standard Model at tree level, from the point of
an Effective Chiral Lagrangian.
We will consider the lowest order
contribution ${\cal O}(g^2)$, ${\cal O}(p^{2})$  plus the
contribution from higher dimensional operators ${\cal O}(g^{4})$,
${\cal O}(g^{2} p^{2})$ and ${\cal O}(p^{4})$
that will explicitly depend on the $\{a_i\}$ coefficients. We shall
{\it not} include the one loop ${\cal O}(g^{4})$, ${\cal O}(g^{2}
p^{2})$ and ${\cal O}(p^{4})$ contributions that, at the same order,
should
be taken into account. This is of course {\it not} quite correct, but it
allows us
to give short closed expressions. Furthermore, this is enough to trace
the
$\log M_H$ dependence in the Standard Model, or the dependence in the
new physics in other models, and argue
the different pros and cons of the several approximations that can be made
when dealing with the Equivalence Theorem.

The exact amplitude for the scattering of four $W_{L}$'s  is obtained
from Fig. 6. The result is
$$\eqalign{A(  W_{L}^{+} W_{L}^{-} \rightarrow W_{L}^{+} W_{L}^{-} )=&
{1 \over 4 M^{4} x^2} \biggl\{ C_{1} (4 M^{2}+ x) \biggl[
4 M^{2} (t^2+tx) + x(t^2 + x^2 + 4tx) \biggr]  \cr
&+ C_{2} \biggl[ 8 M^{4}(2 t^2 + x^2 + 2tx) + 4 M^{2} x (2 t^2 +
x^2 + t x)  \cr
&+ x^{2}(t^2 + x^{2}) \biggr]
+S_{1}\biggl[ -x^{2} {(6 M^{2} + x)}^{2} (2 t + x)\biggr] \cr
&+S_{2}
\biggl[ -16 M^{6} {(2 t +x)}^{2}+ 4
M^{4}(x-2t)(2 (t^2-x^2)-tx)  \cr
&+ 8 M^{2}(t x^3-x t^3) -t^2 x^2(t+2x) \biggr] \cr
&+\sum_{V=\gamma,Z} {1 \over M_{V}^2 -t } \biggl[ A_{V}^{2} \biggl(
-16 M^6 {(2t + x)}^{2}  - 8 M^{2}(t^3 x -x^3 t)  \cr
& + 4 M^{4}(x-2t)(2 (t^2-x^2)-tx)
- t^2 x^2 (t + 2 x)
\biggr) \cr
&+ 2 A_{V} B_{V} t (4 M^{2} + x) \left( 2 M^{2}(2
(t^2-x^2)-tx) + tx(t+2x) \right) \biggr] \cr
&+\sum_{V=\gamma,Z} {1 \over M_{V}^2 -s } \biggl[ A_{V}^{2} \left(
-x^{2}{(6 M^2 + x)}^{2} (2 t+x) \right)  \cr
& + 2 A_{V} B_{V} x^{2}
(4 M^{2} + x) (6 M^{2} + x) (2t+x) \biggr] \biggr\} ,
}\eqno(7.1.1)$$ %
where $C_{1}$ and $C_{2}$ are
defined by
$$\eqalign{&C_{1}=g^{2} \left\{ 1 + g^{2} (a_{4}+a_{8}) - 2
g^{2} ( a_{3} + a_{9} - a_{13} ) \right\}, \cr
&C_{2}=2 g^{4} (a_{4}+a_{5}). }\eqno(7.1.2)$$
The contribution from the self energies of the exchanged vector bosons
is included in the quantities $S_1$ and $S_2$ in (7.1.1)
$$S_{1}=-g^2 \left( c_{w}^{2} {1 \over {(s-M_{Z}^{2})}^{2}}
\Sigma_{Z}(s) + s_{w}^{2} {1 \over {s}^{2} } \Sigma_{\gamma}(s) +
2 s_{w} c_{w} {1 \over s - M_{Z}^{2} } {1 \over s} \Sigma_{\gamma Z}(s)
\right), $$

$$S_{2}=-g^2 \left( c_{w}^{2} {1 \over {(t-M_{Z}^{2})}^{2}}
\Sigma_{Z}(t) + s_{w}^{2} {1 \over {t}^{2} } \Sigma_{\gamma}(t) +
2 s_{w} c_{w} {1 \over t - M_{Z}^{2} } {1 \over t} \Sigma_{\gamma Z}(t)
 \right).
\eqno(7.1.3)$$
The contribution from  $a_{i}$ to the self-energies
involved is given in Appendix C.
Finally $M_V$ stands for the vector boson mass ($V=\gamma,Z$) and
$$\eqalign{
&A_{\gamma}=-i g s_{w} \qquad
 A_{Z}=-i g c_{w} \left( 1 - {1 \over c_{w}^{2}} a_{3}
g^{2} \right) \cr
&B_{\gamma}=-i g^{3} \left( s_{w} a_{3} + s_{w} (a_{1} -
a_{2}) - s_{w} (a_{8}-a_{9}) \right) \cr
&B_{Z}=-i g^{3} \left(- {s_{w}^2 \over c_{w}} a_{3} +
{s_{w}^{2} \over c_{w}} (a_{2}-a_{1}) - c_{w}(a_{8} -
a_{9}) - {1 \over c_{w}} a_{13} \right) }\eqno(7.1.4) $$
Expanding
the exact amplitude (7.1.1)
in inverse
powers of ${v^2}$ one gets
$$\eqalign{A(  W_{L}^{+} W_{L}^{-} \rightarrow W_{L}^{+} W_{L}^{-} )=&
{ 4  \over v^{4} } \biggl[(3 (s^{2}+t^{2})+ 4 s t)(a_{4} +
a_{13}) + 2(s^{2} + t^{2} )(a_{5} - a_{13}) \biggr]
\cr &+ {(s+t) \over v^{2}}  \biggl [ 1 + 6 a_{0} + 6
g^{\prime 2} a_{2} - 2 g^{2} (a_{3} - a_{9} )  \cr
& - g^{2}{(s-2 t)\over s }\left( 12 (a_{4} + a_{13})
+ 8 (a_{5} - a_{13} ) \right) \biggr ] \cr
& -{  g^{2} \over 2 c_{w}^{2} }{1 \over st}(s^{2} + t^{2} + st - 4
t^{2} c_{w}^{2}) + {\cal O}(g^{4}).}\eqno(7.1.5)$$
This result has to be compared with the corresponding scalar amplitude
(Fig. 7 (a)-(e)). The Goldstone boson amplitude at tree level is
$$\eqalign{A(\pi^{+} \pi^{-} \rightarrow \pi^{+} \pi^{-})=&
{ 4  \over v^{4} } \biggl[ (3 (s^{2}+t^{2})+ 4 s t)(a_{4} +
a_{13}) + 2(s^{2} + t^{2} )(a_{5} - a_{13}) \biggr]
\cr &+ {1 \over v^{2}} {(s+ t) } \biggl[ 1 + 6 a_{0}
-g^{2}(12 (a_{4}+a_{13})+8 (a_{5}-a_{13})) \cr
& + 12 g^{2}
s_{w}^{2} (a_{2} - a_{3} - a_{9}) \biggr]
-{2 \over 3} g^{2} - 2 g^{2} s_{w}^{2} {
(s^{2} + t^{2} + s t) \over s t} \cr
& - 8 g^{4} s_{w}^2 (a_{2}-a_{3}-a_{9}) + 4 g^{4} (a_{4} + a_{5})
+ g^{4} v^{2} s_{w}^{2}  {(s + t) \over s t} \cr
&+2 {g^{2} \over v^{2}}{(c_{w}^{2} - s_{w}^{2}) \over c_{w}^{2}}{(s+2 t
- 4 M^{2})  \over M_{Z}^{2} - s} \biggl( c_{w}^{2} (a_{3} + a_{9}) s +
s_{w}^{2} a_{2} s \cr &+ {1 \over 2} v^2 a_{0} + {v^{2} \over 8}
(c_{w}^{2} - s_{w}^{2}) \biggr) \cr
&+2 {g^{2} \over v^{2}}{(c_{w}^{2} - s_{w}^{2}) \over c_{w}^{2}}{(t+2 s
- 4 M^{2})  \over M_{Z}^{2} - t} \biggl( c_{w}^{2} (a_{3} + a_{9}) t +
s_{w}^{2} a_{2} t \cr &+ {1 \over 2} v^2 a_{0} + {v^{2} \over
8} (c_{w}^{2} - s_{w}^{2}) \biggr)
+ {\it self-energies} }\eqno(7.1.6)$$
where `{\it self-energies}' stands for diagrams with self-energy
insertions in the gauge propagators in (a),(b),(d) and (e).
They are of ${\cal O}(g^4)$. Expanding as before
$$\eqalign{A(\pi^{+} \pi^{-} \rightarrow \pi^{+} \pi^{-})=&
{ 4  \over v^{4} } \biggl[ (3 (s^{2}+t^{2})+ 4 s t)(a_{4} +
a_{13}) + 2(s^{2} + t^{2} )(a_{5} - a_{13}) ]
\cr &+ {1 \over v^{2}}(s+t)[ 1 + 6 a_{0} + 6
g^{\prime 2} a_{2} - 6 g^{2} (a_{3} + a_{9} )  \cr
& - g^{2} \left( 12 (a_{4} + a_{13}) + 8
(a_{5} - a_{13} ) \right) \biggr] \cr
& -{ g^{2} \over 2 c_{w}^{2} }{1 \over st}(s^{2} + t^{2} + st
+ {4 \over 3} s t c_{w}^{2}) + {\cal O}(g^{4}).}\eqno(7.1.7)$$
Finally, the contributions with one power of $v^\mu$ lead to
the diagram in Fig. 7 (f).
All the other diagrams entering in this amplitude are of ${\cal O}(g^4)$
and need not be included.
$$\eqalign{
A(\tilde{W}\pi\pi\pi)=
&{8 \over v^{4} x}\left( {\sqrt{s x} - s} \right) \biggl[ ( 6
(s^{2}+t^{2}) + 8 s
t ) (a_{4} + a_{13}) + 4 (s^{2} + t^{2} ) (a_{5} - a_{13}) \biggr]\cr &
+ {4 \over v^{2} x} \left( \sqrt{s x} - s \right) (s + t)
+ {8 \over
v^{2} x} g^{2} \biggl[
 \left( 6 (2 s^{2} +  t^{2}) + 14 st -
\sqrt{s x} (9 s + 7 t) \right) (a_{4}\cr & + a_{13})
+ \left( 4 (2 s^{2} +  t^{2}) + 4 s t -
2 \sqrt{s x} (3 s + t) \right) (a_{5} - a_{13}) +
{\sqrt{s x} \over 2} (s + t) (a_{3} \cr &+ 2 a_{9})
\biggr]
+{ {4} \over 3 x} g^{2} \biggl( 3 (s+t) + 2 x
-12 g^{2} (2 ( s + t) + x) (a_{5}-a_{13}) \cr
&-12 g^{2} (  3 (s + t) + x) (a_{4}+a_{13})
 \biggr)
- 4 g^{2} {\sqrt{s x} \over x} \biggl( 1 + g^{2} (a_{3} - 6 (a_{4}+
a_{13})\cr & - 4 (a_{5} - a_{13}) + 2 a_{9}) \biggr) .
 }\eqno(7.1.8)$$
The expanded amplitude is
$$
A(\tilde{W}\pi\pi\pi)=
{4 g^{2} \over v^{2} }{(s+t) \over s}\biggl[ s a_{3} +6 t
(a_{4} + a_{13})
+ 4 t (a_{5} - a_{13}) + 2 s a_{9} \biggr]
+ {2
} g^{2} \left({1 \over 3}+ {t \over s} \right) + {\cal
O}(g^{4}). \eqno(7.1.9)$$
Adding up eq. (7.1.7) and (7.1.9)
one recovers the result  (7.1.5).
This is a nice example of the verification of the Equivalence Theorem
and we will
use it in a moment to analyze numerically which are the most
relevant pieces that one should
take into account depending on the range of energies. Up to
now the Equivalence Theorem has been mostly considered in
the $g=g^\prime=0$
limit (a small subset of the previous formulae) and found lacking.
The additional terms take good care of the discrepancies.

It is interesting to see that one can easily determine
$a_5^{tree}$ from a comparison between the formulae
derived
in this section and (4.1.6) in the large $M_H$ limit. Expanding the
latter expression in inverse powers of $M_H$ we get
$$\eqalign{
A(  W_{L}^{+} W_{L}^{-} \rightarrow W_{L}^{+} W_{L}^{-} )=&{1 \over v^2}
(s+t) - {g^{2} \over 2 c_{w}^{2}}{1 \over s t} (s^{2} + t^{2} + s t - 4
t^{2} c_{w}^{2}) \cr
&+{1 \over M_{H}^{2}}\left( {1 \over v^2}(s^{2} + t^{2}) + {g^{2} \over
s}(-s^{2} + 2 t^{2} + s t) \right) \cr
&+ {\cal O}(M^{2}/s)
+ {\cal O}({1/ M_{H}^{4}}).
}\eqno(7.1.10)$$
This coincides exactly with the amplitude (7.1.5) if one substitutes
$a_{5}={v^2 / 8 M_{H}^{2}}$ and sets the rest
of $a_{i}$'s equal to zero. The same check can be done for the other
amplitudes.

In processes involving only charged $W$'s there is no dependence on the
two remaining ${\cal O}(p^4)$ operators ${\cal L}_6$ and ${\cal L}_7$.
They appear in processes such as $WW\to ZZ$  involving external $Z$'s.
The values of these coefficients in the minimal Standard Model
are actually best determined by comparing the $S$-matrix elements for these
processes in the Effective Chiral Lagrangian and in the minimal Standard
Model and making use of the Equivalence Theorem itself with $g=g^\prime=0$.
Notice that they give contributions of ${\cal O}(E^4/v^4)$ and thus are
formally unaffected by the additional subleading corrections. (Of
course,
{\it testing experimentally} these ---and the other--- coefficients is
another matter and there one would have to keep the subleading
additional terms and the ${\cal O}(g^{4})$, ${\cal O}(g^{2} p^{2})$
contribution that we have not considered at all.)

\bigskip
\line{\bf 7.2 Domain of Applicability\hfil}
\medskip\noindent
Finding the domain of applicability (or rather the `domain of
usefulness') of the Equivalence Theorem in the framework of the
Effective Chiral Lagrangian is much more subtle than in a renormalizable
theory such as the minimal Standard Model (with a light Higgs).
There is a competition between two type of expansions:
the natural in an effective theory in powers of the energy
(over some scale), and the
expansion in inverse powers of the energy (normalized by some other scale),
 peculiar to the
Equivalence Theorem. The ratio between these two scales and the
number
of terms one takes in each expansion will
determine the window of applicability.

By considering the process described in
detail in section 7.1 we shall try to learn about the
above questions. We shall put
$a_{5}=v^{2} / 8 M_{H}^2$ and $a_{i}=0$ for $i\neq 5$. ${\cal
C}$ is set equal to one. This is
the choice of coefficients that corresponds to treating the
minimal Standard Model at tree level for $E\ll M_H^2$. For us,
however, is just a choice of ${\cal O}(p^4)$ coefficients; the
analysis could well be repeated in the same way for any other
choice.
Our approach will be similar to the one taken in section 4.3. We shall
consider only energies where keeping, at most, the ${\cal O}(p^4)$ terms
in the effective action is meaningful.

We have plotted in Fig 8. the different contributions to the amplitudes
the same three angles as in section 4.3
($\theta=\pi/16, \pi/4, 3\pi/4$). The solid line corresponds to
the exact $A(W^+_L W^-_L \to W^+_LW^-_L)$
amplitude (i.e. to the l.h.s. of
the Equivalence Theorem). The short-dashed line is the Goldstone
boson amplitude with $g=g^\prime=0$ (what is referred in the
text as the `naive' Equivalence Theorem.)
It is clear that this approximation, particularly for small angles,
is quite far from the exact result.

The full Goldstone boson amplitude $A(\pi^+\pi^-\to\pi^+\pi^-)$ with
$g \neq 0$ is
 represented by the
dash-dotted line. The improvement is quite
impressive for most  of the angles quoted. However some small
discrepancies
appear at the $10 \%$  level in the backward direction at low $s$
(case (c)).
In fact in that region setting  $g\neq 0$ worsens the
agreement with the exact amplitude.  For large values of $s$ the
agreement between the dash-dotted line and  the exact result
is actually better than in the Standard Model (Fig. 4). This is because
for such large values of $s$ the Standard Model results are being
very sensitive to the Higgs pole (there we took
 $M_H=1$ TeV), while here in the effective
theory we have obviously no such pole. The cancellation of leading
and
next to leading effects is thus more subtle in the Standard Model that in
the effective theory when $s$ approaches 1 TeV.

Finally we include the additional $A(\tilde{W}\pi\pi\pi)$ amplitude, the
first subleading correction to the Equivalence Theorem. The
$A(\tilde{W}\pi\pi\pi)$ is ${\cal O}(g^2)$ and the result of adding it to the
dash-dotted line
is represented by a long-dashed line, but the reader will not be able
to see it in Fig. 8, except for very small values of $s$ in the
backward direction (c). It just overlaps very nicely with the
exact result almost everywhere.

One could also expand all the contributions in
 powers  of $1/v^2$. This is not a good idea, however. As we see from (a)
the agreement is rather poor (it also was in the Standard Model)
at low values of $s$ and small angles. This is not really a difficulty
of the Equivalence Theorem; if
the $A(W^+_LW^-_L\to W^+_L W_L^-)$ amplitude is also expanded there is
perfect agreement
between the l.h.s. and the r.h.s. of the Equivalence Theorem. Only that
they do not reproduce the exact results. The culprit are the
diagrams with $Z$  exchange at tree level. They are to be kept
without attempting any expansions.

Up to now we have not been very concerned with the fact that
we are dealing with an energy expansion cut at
${\cal O}(E^4)$. The first term that we are throwing away in this
expansion in the effective chiral description of the Standard Model is one
of ${\cal O}(E^6 / v^2 M_H^4)$
 (from tree-level exchange of the Higgs and
assumming that $M_H=1$ TeV).
This  dictates an upper bound to the region
of applicability of the Effective Theory around $E \sim 0.6$ TeV (for this
value of the Higgs mass).
This can be seen
by comparing
Figs. 4 and 8.
This upper bound obviously
depends on the Higgs mass; the higher the Higgs mass  the larger
the region of
coincidence between the Standard Model  and the Effective Chiral
Lagrangian (with the appropriate choice of $a_i$ coefficients, of course).
In any case there is a limiting scale of applicability;
since $4\pi v\simeq 3$ TeV it lies probably around 1.5 TeV.

Are the improvements brought about to the `naive' Equivalence Theorem
necessary to draw physical consequences from experiments? The
answer is obviously positive. In Fig. 9 we have changed the value of
the coefficient $a_5$ from
$v^2/8 M_H^2$, with $M_H=1$ TeV to $N_{TC}/384\pi^2$, with $N_{TC}=16$ (a
popular value in some Technicolor models) and plotted the results for
$\theta=\pi/5$. The figure speaks for itself.

\bigskip
\line{\bf 8. Conclusions\hfil}
\medskip\noindent

In the previous pages we have tried to convey the idea
that the Equivalence Theorem is much more than an easy way of
getting order-of-magnitude estimates for
amplitudes of processes involving longitudinally polarized
$W$'s and $Z$'s. By carefully keeping track of the next-to-leading
corrections
it is possible to compute those amplitudes
in term of other ones involving Goldstone bosons, always evaluated
at $k^2=M^2$, plus
some terms involving just one external $W$ or $Z$ (also evaluated at
$k^2=M^2$),  with
an accuracy that it is good enough to discern different types of
potential `new physics' in the symmetry breaking sector of the Standard
Model. Not only are the calculations technically more convenient
and easy when done with external Goldstone bosons, but
also conceptually clearer, as they are more prone to comparison
with other physical models such as the strong chiral lagrangian.

We have considered the
minimal Standard Model written in the usual linear realization
(the right framework for a light Higgs) and the Effective
Electroweak Chiral lagrangian that encompasses both a heavy Higgs
and other theoretical possibilities in which new physics,
characterized by a scale $\Lambda$,
would creep in through the ${\cal O}(p^4)$ effective operators.
It is a must that we are sensitive to these effective operators,
at least for a range of energies. Otherwise the whole
approach would be useless.

We have seen that for a light Higgs some additional corrections
that we have considered (those involving diagrams with all but one of the
$W$'s replaced by Goldstone bosons) can indeed be neglected at high
energies (but still much lower than $M_H$).
This result ultimately hinges on the perturbative renormalizability
of the model. As soon as we get very close to $M_H$
the additional terms start becoming more relevant, even for $E^2\gg M^2$,
as they are for exceptional momenta configurations in the forward and
backward directions. We have always to keep this in mind.

Other corrections such
as the multiplying factor ${\cal C}$ are clearly necessary if one wants to
work with a one loop precision.
We have for the first time provided all the necessary ingredients to
go beyond tree level by determining the ${\cal C}$ coefficient.
These corrections have been worked out in the usual on-shell scheme. This
is clearly more useful for practical use than other theoretical
possibilities such as non-local gauge fixing terms[8] or
other schemes which are defined in somewhat vage terms[5,10].

In the context of the Effective Chiral Theory, the usual power
counting arguments that have been commomly put forward when
employing
the Equivalence Theorem take a new twist. It should be clearly stated
that the Equivalence Theorem is perfectly valid in the effective
theory. As
far as energy power counting arguments go, the Equivalence Theorem is
both easier and more difficult in the effective theory. For one thing it
is not
always true that the corrections usually lumped under the line ${\cal
O}(M^2/E^2)$ can always be neglected. Because in a non-renormalizable theory
the amplitudes may grow with the energy, these corrections turn out to be
relevant and are required to test the ${\cal O}(p^4)$ terms in the effective
lagrangian.  On the other hand, since the Higgs has been removed from
the spectrum the kinematical singularities that lead to `abnormal'
contributions from the higher order contributions in the $1/E$
expansion are absent.

In addition to being valid, the Equivalence Theorem
remains very useful in an Effective Chiral Lagrangian. It is true that one
must include the
${\cal C}$ factor and the additional $A(\tilde{W}\pi\pi\ldots)$ piece, but
${\cal C}$ in the on-shell scheme depends only on one-loop
self energies and it is finite in the Effective Chiral Lagrangian.
Furthermore the $A(\tilde{W}\pi\pi\ldots)$ additional amplitude needs only
to be computed at tree level since loop corrections would be too small.
Other corrections are completely negligible. In conclusion, the
Equivalence Theorem after being closely scrutinized has been
found sound and well.

\bigskip

\beginsection{\bf Acknowledgements}

We thank J. Cortes, A.Dobado, M.J.Herrero and J.R.Pelaez for
discussions.
We acknowledge the financial support from CICYT grant AEN93-0695 and
CEE grant CHRX CT93 0343. J.M. acknowledges a fellowship from Ministerio
 de Educacion y Ciencia.

\bigskip
\vfill
\eject

\beginsection{\bf Appendix A}

The renormalized self-energies of the Standard Model, that we will use
in the Equivalence Theorem are expressed in
terms of the bare self-energies in an arbitrary gauge in the following
way

$$\eqalign{{\hat \Sigma}_{w}(k^2)&=\Sigma_{w}(k^2)-\xi_{2 }
\Sigma_{T }(M^{2})+\delta Z_{w} (k^2- \xi_{2}
M^{2})-\xi_{2} M^{2} \delta Z_{\xi_{2}}, \cr
{\hat \Sigma}_{T}(k^2)&=\Sigma_{T }(k^2)-
\Sigma_{T }(M^{2})+\delta Z_{W} (k^2-
M^{2}),\cr
\xi_{1}{\hat \Sigma}_{L}(k^2)&=\xi_{1} \Sigma_{L
}(k^2)- \xi_{1}\Sigma_{T }(M^2)+\delta Z_{W}
(k^2-\xi_{1} M^{2})-k^2 \delta
Z_{\xi_{1}}, \cr
{\hat \Sigma}_{W w}(k^2)&=\Sigma_{W w}(k^2)+ {1 \over 2} M
(\delta Z_{\xi_{1}} - \delta Z_{\xi_{2}})
.}
\eqno(A.1) $$
In the previous expression ${\it M}$ is the renormalized ${\it W}$-boson
mass.
In the on-shell scheme the renormalization constants can be written in
terms of bare self-energies
$$\eqalign{&\delta Z_{M}={1 \over M^{2}} \Sigma_{T }(M^2), \cr
           &\delta Z_{\xi_{1}}={1 \over M^{2}}
\left( \Sigma_{L
}(\xi_{1} M^{2})-\Sigma_{T }(M^{2}) \right),
\cr
           &\delta Z_{\xi_{2}}={1 \over M^{2}}
\left({ \Sigma_{w}( \xi_{2} M^{2})\over
\xi_{2}}-\Sigma_{T }(M^{2})\right).}\eqno(A.2)
$$
The last two relations are deduced from eq.(A.1) by imposing the
on-shell condition
$$\eqalign{&{\hat \Sigma}_{w}(\xi_{2} M^2)=0, \cr
           &{\hat \Sigma}_{L}(\xi_{1} M^2)=0.}\eqno(A.3)
$$
The external renormalization constants for the $W$ and $\pi$ fields are
given by
$$\eqalign{
&{\tilde Z}_{W}=1-
{\partial \Sigma_{T}(k^{2}) \over \partial k^2}\vert_{k^{2}=M^{2}} \cr
&{\tilde Z}_{w}=1-
{\partial \Sigma_{w}(k^{2}) \over \partial
k^2}\vert_{k^{2}=\xi_{2} M^{2}} }\eqno(A.4)$$
They differ from the field renormalization constants $Z_{W},Z_{w}$
by a finite amount.

Here we will give the
divergent parts and $M_H$ dependence of the bare
self-energies that are
needed in order to evaluate the ${\cal C}$ factor in the Standard
Model. In these expressions we have taken
$\xi_{1}=\xi_{2}=\xi$. The divergent contributions are
$$\eqalign{
&\Sigma_{T}(k^2)={g^2 \over 8 \pi^2}{1 \over \epsilon} M^2 (
+{3 \over 2} - {3 \over 4 c_{w}^2} - \xi ({1 \over 4 c_{w}^2} -{1
\over 2}))-{g^2 \over 8 \pi^2}{k^2 \over \epsilon} (-{25 \over 6} +
\xi), \cr
&\Sigma_{L }(k^2)={g^2 \over 8 \pi^2}{1 \over \epsilon} M^2 (
+{3 \over 2} - {3 \over 4 c_{w}^2} - \xi ({1 \over 4 c_{w}^2} -{1
\over 2})), \cr
 &\Sigma_{w}(k^2)={g^2 \over 8 \pi^2}{1 \over \epsilon} k^{2} ({3\over
2}+{3 \over 4 c_{w}^2}-\xi ({1 \over 2} + {1 \over 4 c_{w}^2})).}
\eqno(A.6)$$
And the Higgs dependent contribution is
$$\eqalign{
&\Sigma_{T}(k^2)=
{g^2 \over 8 \pi^2} M^{2}(-{1 \over 16}
{M_{H}^2 \over M^2} - {3 \over 8} \ln{M_{H}^{2}} )-{g^2 \over 8
\pi^2} {k^2 \over 24} \ln{M_{H}^2}, \cr
&\Sigma_{L}(k^2)=
{g^2 \over 8 \pi^2} M^{2}(-{1 \over 16}
{M_{H}^2 \over M^2} - {3 \over 8} \ln{M_{H}^{2}} ),
\cr
 &\Sigma_{w}(k^2)=
{g^2 \over 8 \pi^2} k^2 ({1 \over 16}
{M_{H}^2 \over M^2} + ({3 \over 8}-{\xi \over 4}) \ln
{M_{H}^{2}} ). }\eqno(A.7) $$

\bigskip

\beginsection{\bf Appendix B}

In the on-shell scheme one usually imposes a unit residue wave function
renormalization condition on the Higgs self-energy.
In the effective field theory the Higgs has dissappeared and it does
not seem very sensible to give
renormalization conditions over a non-existing field.
We will replace the condition from the Higgs self-energy
to the Goldstone boson one.
We will follow the discussion of renormalization conditions in
the section 4 of the paper by Bohm et al[16].
The renormalization conditions in the on-shell scheme are:

\noindent (1) The propagators have poles at the physical masses
of the particles

$$ \eqalign{{\hat \Sigma}_{T}^{W}(M^{2})=0 \qquad {\hat
\Sigma}_{T}^{Z}(M_{Z}^{2})=0 \qquad
{\hat \Sigma}_{H}(M_{H}^2)=0} \eqno(B.1) $$

\noindent (2) The electric charge is defined as in QED implying

$$ {\hat \Sigma}_{T}^{\gamma Z}(0)=0 \eqno(B.2)$$

\noindent (3) The photon and Higgs propagator have unit residue
$$ \eqalign{{1 \over k^2}{\hat
\Sigma}_{T}^{\gamma}(k^{2})\vert_{k^{2}=0}=0 \qquad
{ \partial \over \partial k^{2}} {\hat
\Sigma}_{H}(k^{2})\vert_{k^{2}= M_{H}^{2} }=0} \eqno (B.3) $$

\noindent (4) Similar requirements are imposed on the unphysical
sector

$$ {\hat \Sigma}_{L}^{W}(\xi^{W} M^{2})=0 \qquad {\hat
\Sigma}_{L}^{Z}(\xi^{Z} M_{Z}^{2})=0 \qquad {\hat
\Sigma}_{w^{+}}(\xi^{W} M^{2})=0 \qquad {\hat
\Sigma}_{w^{3}}(\xi^{Z} M_{Z}^{2})=0  $$
$$ {\hat \Sigma}^{\gamma w^{3}}(0)=0 \qquad {1 \over k^{2}}
{\hat \Sigma}_{L}^{\gamma}(k^2) \vert_{k^{2}=0}=0
\eqno(B.4) $$

\noindent (5) The vanishing tadpole condition

$$v^{2} + {\mu^{2} \over \lambda}+ \delta T=0 \eqno(B.5) $$
where $\delta T$ is the tadpole contribution generated in perturbation
theory. In the Effective Chiral Lagrangian
the requirements on the Higgs self-energy
and the tadpole condition do not make sense anymore.
On the other hand we have two renormalization constants less, namely
$\delta \lambda$ and $\delta \mu$.
When we use the Effective Chiral
Lagrangian we replace $\omega \to \pi$ and instead of the second
equation of (B.3) we take
$$ \eqalign{ { \partial
\over \partial k^{2}} {\hat \Sigma}_{\pi^{+}}(k^{2})\vert_{k^{2}=\xi^{W}
M^2 } = 0 }. \eqno (B.6) $$
This fixes $Z_{\pi}$ and
we are still left with a
compatible system of equations.
By solving the remaining equations it is quite easy to see that the
change in $Z_{\pi}$ affects only to $\delta v$.
In order to evaluate these changes one can use the
first condition of eq. (B.1)
$$\eqalign{&{\delta v \over v}={1 \over 2} \left(
-{\Sigma_{T}^{W}(M^{2}) \over M^{2}} - \delta Z_{W} - 2 {\delta
g \over g} + {\delta Z_{\pi}} \right). }\eqno (B.7) $$

\bigskip

\beginsection{\bf Appendix C}

The set of ${\it C}$ and ${\it P}$ and $SU(2)_{L} \times U(1)$ {\cal gauge}
invariant operators ${\cal L}_{i}$ are

$$\vcenter{\openup1\jot \halign{
$\hfil#$&${}#\hfil$&\qquad$\hfil#$&${}#\hfil$    \cr
{\cal L}&_{0}={1 \over 4} a_{0} v^2  {T}_{\mu} {T}^{\mu}
&{\cal L}&_{1}={1 \over 2} a_{1} g g' B_{\mu\nu} {\rm Tr} T W^{\mu \nu}
\cr \cr
{\cal L}&_{2}={i } a_{2} g' B_{\mu\nu} {\rm Tr}[{T} V^{\mu} V^{\nu}]
&{\cal L}&_{3}=-i a_{3} g {\rm Tr}[W^{\mu\nu}[V_{\mu},V_{\nu}]]
\cr \cr
{\cal L}&_{4}=a_{4} {\rm Tr}[ V_{\mu} V_{\nu}] {\rm Tr}[ V^{\mu}
V^{\nu} ]
&{\cal L}&_{5}=a_{5} {\rm Tr}[ V^{\mu} V_{\mu}] {\rm Tr}[ V^{
\nu} V_{\nu}] \cr \cr
{\cal L}&_{6}=a_{6} {\rm Tr}[ V_{\mu} V_{\nu} ] T^{ \mu}
T^{\nu}
&{\cal L}&_{7}=a_{7} {\rm Tr}[ V_{\mu} V^{\mu}] T^{ \nu}
{T}_{\nu} \cr \cr
{\cal L}&_{8}=-{1 \over 4} a_{8}{g^2 } {\rm Tr}[ T W_{\mu \nu} ]
{\rm Tr}[ T W^{\mu \nu} ]
&{\cal L}&_{9}=-{i} a_{9} {g} {\rm Tr}[ T W^{ \mu \nu}]
{\rm Tr}[ T V^{\mu} V^{\nu} ]  \cr \cr
{\cal L}&_{10}=a_{10} ({T}_{\mu} {T}_{\nu})^2
&{\cal L}&_{11}=a_{11} {\rm Tr}[({\cal D}_{\mu} V^{\mu})^2] \cr
 \cr
{\cal L}&_{12}=a_{12} {\rm Tr}[T{\cal D}_{\mu} {\cal D}_{\nu} V^{\nu}]
{T} ^{\mu}
&{\cal L}&_{13}={1 \over 2} a_{13} ({\rm Tr}[T {\cal
D}_{\mu}V_{\nu}])^2 \cr \cr
}}
\eqno(C.1)$$
where
$$ \eqalign{
&V_{\mu}= (D_{\mu}U)U^{\dag}\qquad  T=U \tau_{3} U^{\dag} \qquad
{T}_{\mu}= {\rm Tr} T V_{\mu}\qquad  \cr \cr
&{\cal D}_{\mu}
O(x)=\partial_{\mu} O(x) + i g \left[ W_{\mu},O(x) \right ]}\eqno(C.2)
$$
Expanding the operators of eq.(C.1) up to two fields
one finds
$$\eqalign{ \Sigma_{T }^{W}(k^2)=&0 \qquad
\Sigma_{L }^{W}(k^2)=-g^2 a_{11} k^{2}
\qquad
\Sigma_{\pi}(k^2)={4  a_{11} } {1 \over v^2} k^4  \cr
 \Sigma_{T}^{Z}(k^{2})=&k^{2} ( c_{w}^{2} g^{2} a_{8}
+ 2 s_{w}^{2}
g^{2} a_{1} + {g^{2} \over c_{w}^{2}} a_{13} )
+2 M_{Z}^{2} a_{0} \cr
 \Sigma_{T}^{\gamma}(k^{2})=&k^2 ( s_{w}^{2} g^{2} (a_{8} - 2
a_{1}) ) \cr
\Sigma_{T}^{\gamma Z}(k^{2})=&k^{2} \left( s_{w} c_{w} g^{2} a_{8} -
(c_{w}^{2} - s_{w}^{2}) g g^{\prime} a_{1} \right) }\eqno(C.3)
$$

\beginsection{\bf Appendix D}

\def\doble {\buildrel \leftrightarrow \over \partial}
We expand the Standard Model lagrangian in the non-linear variables
in the number of fields. The hats mean that the two, three and four
$\hat {\cal L}$ include the contributions from the $L_{GF}$ that do not
depen on $\xi$ and in $\hat {\cal L}_{GF}$ it is written the $\xi$
dependent pieces.

$$ {\cal L}=\hat{\cal L}_{2} + \hat{\cal L}_{3} + \dots +
\hat{\cal L}_{GF\, 2}+ \hat {\cal  L}_{GF\, 3}+\dots
$$
$$\eqalign{\hat {\cal {L}}_{2}=& {1 \over 2} { \partial_{\mu} \rho}
{\partial_{\mu} \rho}+
{1 \over 2}{\partial_{\mu}
\pi^{3} \partial_{\mu} \pi^{3}}
+ { \partial_{\mu} \pi^{+}}\,{\partial_{\mu} \pi^{-}}
 - v \rho (\mu^2+\lambda v^2)
-{1 \over 2} \rho^2 (\mu^2+3 v^2 \lambda)
 \cr
& + {1 \over 4 s_{w}^2} {e}^2\,{v^2}\,{
W_{\mu}^{+}}\,{ W_{\mu}^{-}}
 + {1 \over 8 s_{w}^2 c_{w}^2} {e}^2\,{v^2}\,
{{Z_{\mu}}^2} \cr
{\hat {\cal L}}_{3}=&- \lambda v \rho^3
+{1 \over v}{ \rho}\,{ \partial_{\mu} \pi^{3}}\,{\partial_{\mu} \pi^{3}}
+ {2 \over v}{ \rho}\,{ \partial_{\mu} \pi^{+}}\,{ \partial_{\mu}
\pi^{-}}
+ { e \over 2 s_{w}}\,(\,{\rho}\,{\doble_{\mu}
\pi^{-}}\,)\,{
W_{\mu}^{+}} \cr
&+ { e \over
2 s_{w}}\,(\,{\rho}\,{\doble_{\mu}
\pi^{+}}\,)\,{ W_{\mu}^{-}}
- {i\over 2 s_{w}}\,{ e}\,(\,\pi^{3}\,{\doble_{\mu}
\pi^{-}}\,)\,{ W_{\mu}^{+}}
+ {i\over 2 s_{w}}\,{ e}\,(\,\pi^{3}\,{
\doble_{\mu} \pi^{+}}\,)\,{ W_{\mu}^{-}}  \cr
& + i\,{ e}\,A_{\mu}\,(\,\pi^{+}\,{\doble_{\mu} \pi^{-}}\,)
+ { e \over 2 s_{w} c_{w}}\,
(\,\rho\,{\doble_{\mu} \pi^{3}}\,)\,
Z_{\mu}
 - {i\over 2}\,{ e}\,{c_{w}^2-s_{w}^2 \over s_{w} c_{w}}
\,(\,\pi^{-}\,{ \doble_{\mu}
\pi^{+}}\,)\, Z_{\mu} \cr
 &+ {1\over 2 s_{w}^2}{{{{ e}}^2}\,{ \rho}\,v\,{
W_{\mu}^{+}}\, { W_{\mu}^{-}}}
+{1 \over 4 s_{w}^2 c_{w}^2}{{{{ e}}^2}\,{
\rho}\,v\,
{{Z_{\mu}}^2}}
+ {i\over 2 c_{w}}\,{{{
e}}^2}\,v\,\pi^{-}\,{ W_{\mu}^{+}}\,
Z_{\mu} - {i\over
2 c_{w}}\,{{{ e}}^2}\,v\,\pi^{+}\,{ W_{\mu}^{-}}\,
Z_{\mu} \cr
 & - {i\over 2 s_{w}}\,{{{
e}}^2}\,v\,\pi^{-}\,{ W_{\mu}^{+}}\,A_{\mu}
+ {i\over 2 s_{w}}\,{{{
e}}^2}\,v\,\pi^{+}\,{
W_{\mu}^{-}}\,A_{\mu}
\cr
 }$$
$$\eqalign{{\hat {\cal L}}_{4}=&-{1 \over 4} \lambda \rho^4 + {1\over 2
v^2} {{{{ \rho}}^2}\,{{{ \partial_{\mu}
\pi^{3}}}{\partial_{\mu} \pi^{3}}}} +{1 \over v^2} {{{{ \rho}}^2}\,{
\partial_{\mu} \pi^{+}}\,{ \partial_{\mu} \pi^{-}}}
 +{1\over 3 v^2}
{\partial_{\mu}  \pi^{3}}\, \pi^{3}
 ({ \partial_{\mu} \pi^{+}}\,\pi^{-}\,
 +{ \partial_{\mu} \pi^{-}}\,\pi^{+}\,) \cr
& -{1 \over 3 v^2}({ \partial_{\mu}
\pi^{+}}  { \partial_{\mu} \pi^{-}}(
 \pi^{+}  \pi^{-}+
{{(\pi^{3})}^2})
+ { \partial_{\mu}
\pi^{3}}{\partial_{\mu} \pi^{3}} \pi^{+}  \pi^{-})
 + {1 \over 6 v^2}{ \partial_{\mu}
\pi^{+}}{\partial_{\mu} \pi^{+}} {(\pi^{-})}^2 \cr
&+ {1 \over 6 v^2}{ \partial_{\mu}
\pi^{-}}{\partial_{\mu} \pi^{-}}\,({\pi^{+}})^2
 + {1 \over 2 v s_{w}}{{ e}\,{{{ \rho}}^2}\ ({
\partial_{\mu} \pi^{-}}\,{ W_{\mu}^{+}}}
 + { \partial_{\mu} \pi^{+}}\, {W_{\mu}^{-}}
  +{1 \over c_{w}}{ \partial_{\mu} \pi^{3}}\,Z_{\mu})
\cr
 &
+{2 \over v} {i\,{ e}\,{ \rho}\,A_{\mu}\,(\,\pi^{+}\,{ \doble_{\mu}
\pi^{-}}})
-{i \over v}
{ e}\,{ \rho}\,{c_{w}^2-s_{w}^2 \over s_{w} c_{w}}\,
(\,\pi^{-}\,{
\doble_{\mu} \pi^{+}}\,)\,Z_{\mu} \cr
&+{1 \over 4 v s_{w}}{e}\,(\, 2\,({\pi^{-}})^{2}\,
{\partial_{\mu} \pi^{+}}\,
+ 4 {i}\,{ \rho}\,(\,\pi^{-}\,{
\doble_{\mu} \pi^{3}}\,)\,
+\pi^{-}\,{ \doble_{\mu}
{{(\pi^{3})}^2}}\,)\, {W_{\mu}^{+}} \cr
&+{1 \over 4 v s_{w}}{e}\,(\, 2\,({\pi^{+}})^{2}\,
{\partial_{\mu} \pi^{-}}\,
- 4 {i}\,{ \rho}\,(\,\pi^{+}\,{
\doble_{\mu} \pi^{3}}\,)\,
+\pi^{+}\,{ \doble_{\mu}
{{(\pi^{3})}^2}}\,)\, {W_{\mu}^{-}} \cr
&+{1 \over 2 v s_{w} c_{w}}
{e}\,(\,\pi^{-}\,(
{\pi^{3} \doble_{\mu} \pi^{+}}) + {\pi^{3} \partial_{\mu} \pi^{-}
\pi^{+}}+ {1 \over 2}{ {(\pi^{3})}^2 \partial_{\mu} \pi^{3} }\,)\,
Z_{\mu} \cr
&+{1 \over 4 s_{w}^2} {{{{
e}}^2}\,{{{ \rho}}^2}\, { W_{\mu}^{+}}\,{
W_{\mu}^{-}}}
 + {1 \over 8 s_{w}^2 c_{w}^2}{{{{ e}}^2}\,{{{
\rho}}^2}\,
{{Z_{\mu}}^2}}
- {{{ e}}^2}\,\pi^{+}\,\pi^{-}\,{{Z_{\mu}}^2}
+ {{{e}}^2}\,\pi^{+}\,\pi^{-}\,{{A_{\mu}}^2} \cr
&+{1 \over 2 } { e}^2\,(\,{\pi^{3}+2 i \rho}\,)\,\pi^{-}\,
(\,{1 \over c_{w}} Z_{\mu}\,W_{\mu}^{+}
-{1 \over s_{w}}\,A_{\mu}\,W_{\mu}^{+}\,)\
 +{e}^2\,{c_{w}^2-s_{w}^2 \over s_{w} c_{w}}\,
\pi^{+}\,\pi^{-}\, Z_{\mu}\,A_{\mu} \cr
&+{1 \over 2 } { e}^2\,(\,{\pi^{3}-2 i \rho}\,)\,\pi^{+}\,
(\,{1 \over c_{w}} Z_{\mu}\,W_{\mu}^{-}
-{1 \over s_{w}}\,A_{\mu}\,W_{\mu}^{-}\,)\
}$$

$$\eqalign{{\hat {\cal  L}}_{GF\, 2}=
&-{1 \over 2}({ c_{w}^2 \over \xi^{B}}+{s_{w}^2 \over  \xi^{W}})\,
{( \partial_{\mu}A_{\mu})}^2
-{1 \over 2}({ c_{w}^2 \over  \xi^{W}}+{s_{w}^2 \over  \xi^{B}})\,
{( \partial_{\mu}Z_{\mu})}^2
+s_{w} c_{w}\, ({1\over \xi^{B}}-{1\over
\xi^{W}})\,\partial_{\mu}A_{\mu}\, \partial_{\nu}Z_{\nu} \cr
&-{ 1 \over {{ \xi^{W}}}}\,
{ \partial_{\mu}
W_{\mu}^{+}}\,{ \partial_{\nu} W_{\nu}^{-}}
 -{v^2 \over 8}\, (g^2 \xi^{W} +g^{\prime 2} \xi^{B})\, (\pi^3)^2
-{v^2 \over 4}\, g^2 \xi^{W}\, \pi^{+} \pi^{-}\cr
{\hat {\cal L}}_{GF\, 3}=&-{v \over 4}\, (g^2 \xi^{W}+
g^{\prime 2} \xi^{B})\,\rho\, {(\pi^3)}^2 - {v \over 2}\, g^2 \xi^{W}
\,\rho\, \pi^{+} \pi^{-} \cr
{\hat {\cal L}}_{GF\, 4}=&{1 \over 24}\,(g^2 \xi^{W}+
g^{\prime 2} \xi^{B})\,{(\pi^3)}^4+{1 \over 12}\,
(2 g^2 \xi^{W}+g^{\prime 2} \xi^{B}) \, (\pi^3)^2\,
\pi^{+}\pi^{-} \cr
&+{g^2 \over 6} \xi^{W}\, (\pi^{+})^2
(\pi^{-})^2
-{1 \over 8} (g^2 \xi^{W} + g^{\prime 2} \xi^{B}) \rho^2 (\pi^3)^2
-{g^2 \over 4} \xi^{W} \rho^2 \pi^{+} \pi^{-}
\cr} $$
\medskip
The lagrangian of the Standard Model in the non-linear realization, but
now in the linear
gauge (2.2.12) is the following (with
 ${\hat {\cal L}}_{GF\, 3,4}=0$)
$$\eqalign{
{\hat {\cal L}}_{3}=&- \lambda v \rho^3
+{1 \over v}{ \rho}\,{ \partial_{\mu} \pi^{3}}\,{\partial_{\mu} \pi^{3}}
+ {2 \over v}{ \rho}\,{ \partial_{\mu} \pi^{+}}\,{ \partial_{\mu}
\pi^{-}}
+ { e \over s_{w}}\,(\,{\rho}\,{\partial_{\mu}
\pi^{-}}\,)\,{
W_{\mu}^{+}}
\cr &+ { e \over
s_{w}}\,(\,{\rho}\,{\partial_{\mu}
\pi^{+}}\,)\,{ W_{\mu}^{-}}
- {i\over 2 s_{w}}\,{ e}\,(\,\pi^{3}\,{\doble_{\mu}
\pi^{-}}\,)\,{ W_{\mu}^{+}}
+ {i\over 2 s_{w}}\,{ e}\,(\,\pi^{3}\,{
\doble_{\mu} \pi^{+}}\,)\,{ W_{\mu}^{-}}  \cr
& + i\,{ e}\,A_{\mu}\,(\,\pi^{+}\,{\doble_{\mu} \pi^{-}}\,)
+ { e \over  s_{w} c_{w}}\,
(\,\rho\,{\partial_{\mu} \pi^{3}}\,)\,
Z_{\mu}
 - {i\over 2}\,{ e}\,{c_{w}^2-s_{w}^2 \over s_{w} c_{w}}
\,(\,\pi^{-}\,{ \doble_{\mu}
\pi^{+}}\,)\, Z_{\mu} \cr
&+ {1\over 2 s_{w}^2}{{{{ e}}^2}\,{ \rho}\,v\,{
W_{\mu}^{+}}\, { W_{\mu}^{-}}}
+{1 \over 4 s_{w}^2 c_{w}^2}{{{{ e}}^2}\,{
\rho}\,v\,
{{Z_{\mu}}^2}}
+ {i\over 2 c_{w}}\,{{{
e}}^2}\,v\,\pi^{-}\,{ W_{\mu}^{+}}\,
Z_{\mu} - {i\over
2 c_{w}}\,{{{ e}}^2}\,v\,\pi^{+}\,{ W_{\mu}^{-}}\,
Z_{\mu} \cr
& - {i\over 2 s_{w}}\,{{{
e}}^2}\,v\,\pi^{-}\,{ W_{\mu}^{+}}\,A_{\mu}
+ {i\over 2 s_{w}}\,{{{
e}}^2}\,v\,\pi^{+}\,{
W_{\mu}^{-}}\,A_{\mu}
\cr
 }$$
$$\eqalign{{\hat {\cal L}}_{4}=&-{1 \over 4} \lambda \rho^4 +
{1\over 2 v^2} {{{{ \rho}}^2}\,{{{ \partial_{\mu}
\pi^{3}}}{\partial_{\mu} \pi^{3}}}} +{1 \over v^2} {{{{ \rho}}^2}\,{
\partial_{\mu} \pi^{+}}\,{ \partial_{\mu} \pi^{-}}}
 +{1\over 3 v^2}
{\partial_{\mu}  \pi^{3}}\, \pi^{3}
 ({ \partial_{\mu} \pi^{+}}\,\pi^{-}\,
 +{ \partial_{\mu} \pi^{-}}\,\pi^{+}\,) \cr
& -{1 \over 3 v^2}({ \partial_{\mu}
\pi^{+}}  { \partial_{\mu} \pi^{-}}(
 \pi^{+}  \pi^{-}+
{{(\pi^{3})}^2})
+ { \partial_{\mu}
\pi^{3}}{\partial_{\mu} \pi^{3}} \pi^{+}  \pi^{-})
 + {1 \over 6 v^2}{ \partial_{\mu}
\pi^{+}}{\partial_{\mu} \pi^{+}} {(\pi^{-})}^2 \cr
&+ {1 \over 6 v^2}{ \partial_{\mu}
\pi^{-}}{\partial_{\mu} \pi^{-}}\,({\pi^{+}})^2
 + {1 \over 2 v s_{w}}{{ e}\,{{{ \rho}}^2}\ ({
\partial_{\mu} \pi^{-}}\,{ W_{\mu}^{+}}}
 + { \partial_{\mu} \pi^{+}}\, {W_{\mu}^{-}}
  +{1 \over c_{w}}{ \partial_{\mu} \pi^{3}}\,Z_{\mu})
\cr
&+{2 \over v} {i\,{ e}\,{ \rho}\,A_{\mu}\,(\,\pi^{+}\,{ \doble_{\mu}
\pi^{-}}})
-{i \over v}
{ e}\,{ \rho}\,{c_{w}^2-s_{w}^2 \over s_{w} c_{w}}\,
(\,\pi^{-}\,{
\doble_{\mu} \pi^{+}}\,)\,Z_{\mu}
+{1 \over 3 v s_{w}}{e}\,(\,
\pi^{-} (\,\pi^{-}\,{ \doble_{\mu}
{{\pi^{+}}}}\,)\, \cr
&+({\pi^{3}} + 3 i {\rho} ) (\,\pi^{-}\,{ \doble_{\mu}
{{\pi^{3}}}}\,)\
)\,{ W_{\mu}^{+}}
+
{1 \over 3 v s_{w}}{e}\,(\,
\pi^{+} (\ \pi^{+}\,{ \doble_{\mu}
{{\pi^{-}}}}\,)\, \cr
&+(\pi^3 - 3 i { \rho})\,(\,\pi^{+}\,{
\doble_{\mu} \pi^{3}}\,)\,
)\ { W_{\mu}^{-}}
+{1 \over 3 v s_{w} c_{w}}
{e} (\,
\pi^{-}\,({\pi^{3} \doble_{\mu} \pi^{+}}) +
\pi^{+}\,({\pi^{3} \doble_{\mu} \pi^{-}})
)\, {Z_{\mu}} \cr
&+{1 \over 4 s_{w}^2} {{{{
e}}^2}\,{{{ \rho}}^2}\, { W_{\mu}^{+}}\,{
W_{\mu}^{-}}}
 + {1 \over 8 s_{w}^2 c_{w}^2}{{{{ e}}^2}\,{{{
\rho}}^2}\,
{{Z_{\mu}}^2}}
- {{{ e}}^2}\,\pi^{+}\,\pi^{-}\,{{Z_{\mu}}^2}
+ {{{e}}^2}\,\pi^{+}\,\pi^{-}\,{{A_{\mu}}^2} \cr
&+{1 \over 2 } { e}^2\,(\,{\pi^{3}+2 i \rho}\,)\,\pi^{-}\,
(\,{1 \over c_{w}} Z_{\mu}\,W_{\mu}^{+}
-{1 \over s_{w}}\,A_{\mu}\,W_{\mu}^{+}\,)\
 +{e}^2\,{c_{w}^2-s_{w}^2 \over s_{w} c_{w}}\,
\pi^{+}\,\pi^{-}\, Z_{\mu}\,A_{\mu} \cr
&+{1 \over 2 } { e}^2\,(\,{\pi^{3}-2 i \rho}\,)\,\pi^{+}\,
(\,{1 \over c_{w}} Z_{\mu}\,W_{\mu}^{-}
-{1 \over s_{w}}\,A_{\mu}\,W_{\mu}^{-}\,)\
}$$

\beginsection{\bf References}

\item{[1]}{ J.M. Cornwall, D.N. Levin and G.Tiktopoulos, Phys. Rev.
D10 (1974) 1145 ; B.W. Lee, C. Quigg and H. Thacker, Phys. Rev. D16 (1977)
1519; G.J. Gounaris, R. Kogerler and H. Neufeld, Phys. Rev. D34 (1986)
3257}

\item{[2]}{ M.S. Chanowitz and M.K. Gaillard, Nucl. Phys. B261 (1985)
379}

\item{[3]}{ A.Dobado and J.R.Pelaez, Nucl.Phys. B425 (1994) 110;
Phys. Lett. B329 (1994) 469}

\item{[4]}{ C. Grosse-Knetter and I.Kuss, Bielefeld preprint
BI-TP 94/10, March 1994; C. Grosse-Knetter, Bielefeld preprint
BI-TP 94/25 May 1994}

\item{[5]}{ H.J.He,Y.P.Kuang and X.Li, Phys. Lett. B329 (1994) 278}

\item{[6]}{ T.Appelquist and C.Bernard, Phys. Rev. D22 (1980) 200;
A.Longhitano, Phys. Rev. D22 (1980) 1166;
A. Longhitano, Nucl. Phys. B188 (1981) 118}

\item{[7]}{ J.Gasser and H.Leutwyler, Ann. of Phys. 158 (1984) 142,
Nucl. Phys. B250(1985) 465 and 517}

\item{[8]}{ Y.P.Yao and C.P.Yuan, Phys.Rev. D38 (1988) 2237}

\item{[9]}{ J.Bagger and C.Schmidt, Phys. Rev. D41 (1990) 264}

\item{[10]}{ H.J.He,Y.P.Kuang and X.Li, Phys. Rev. Lett. 69
(1992) 2619 ; Phys.Rev. D49 (1994) 4842}

\item{[11]}{ P.B.Pal, University of Texas preprint, DOE-ER-40757-045,
Mar. 1994}

\item{[12]}{ H.J.He,Y.P.Kuang and C.P. Yuang, preprint
VPI-IHEP-94-04 Sep 1994}

\item{[13]}{ A.Dobado, J.R.Pelaez and M.T. Urdiales, ``The
applicability of the Equivalence Theorem in Chiral Perturbation Theory'',
Contributed
paper to the 27${\sl th}$ International Conference in High Energy
Physics 1994}

\item{[14]}{ M.Veltman, Acta Phys. Pol. B8 (1977), 475}

\item{[15]}{ W.J. Marciano and A. Sirlin, Phys.Rev. D22 (1980) 2695;
K.I. Aoki, Z. Hioki, R. Kawabe, M. Konuma and T. Muta, Suppl. of
the Progress of Theoretical Physics 73 (1982) 1;
J. Fleischer and F. Jegerlehner, Nucl.Phys. B 228 (1983) 1; G.Burgers
and W.Hollik, in Polarization at LEP, CERN
Yellow Report,ed. G. Alexander et al. (CERN, Geneva, 1988); M.Consoli
and W.Hollik, in Z Physics at LEP1, CERN Yellow Report, ed. G.Altarelli
et al. (CERN, Geneva 1989) G. Burgers and F.Jegerlehner, ibid}

\item{[16]}{ M.Bohm,H.Spiesberger and W.Hollik, Fortschr. Phys. 34
(1986) 687}

\item{[17]}{ F.Feruglio in Lectures at the $2^{nd}$ NATO Seminar,
Parma, Univ. di Padova (1992) DFPD92-TH-/50}

\item{[18]}{ M.J.Herrero and E.R.Morales, Nucl.Phys. B418 (1994) 431}

\item{[19]}{ D.Espriu and J.Matias, preprint Univ. of Barcelona,
UB-ECM-PF-94/12 June 1994 (to appear in Phys.Lett. B)}

\item{[20]}{ J. Donoghue, E. Golowich and B. Holstein, in {\sl
Dynamics of The Standard Model}, Cambridge University Press.}

\item{[21]}{ R.Haag , Phys. Rev. 112 (1958) 669; S.Coleman, J.Wess and
B.Zumino, Phys.Rev.177 (1969) 2239;C.G.Callan, S Coleman,
J.Wess and B.Zumino, Phys.Rev. 177 (1969) 2247}

\item{[22]}{ J.C. Taylor,{\sl Gauge Theories of Weak Interactions},
Cambridge University Press.}

\item{[23]}{ J.Bagger, S.Dawson and G.Valencia, Nucl. Phys. B399 (1993)
364;

\item{[24]}{ H.Veltman, Phys. Rev. D41 (1990) 2294}

\item{[25]}{ G. Barton, {\sl Introduction to Advanced Field Theory}, J.
Wiley}

\item{[26]}{ W. Hollik, DESY preprint 88-188;
 B.W.Lynn, R.G. Stuart, Nucl. Phys. B253 (1985) 216}

\item{[27]}{ M.J. Duncan, G.L. Kane and W.W. Repko, Nucl. Phys. B272
(1986) 517; S. Dawson and S. Willenbrock, Phys. Rev. Lett. 62 (1989)
1232}

\item{[28]}{ V.Barger, K.Cheung, T.Han,
R.J.N. Phillips, Phys.Rev. D42 (1990) 3052}

\item{[29]}{ R. Renken and M.Peskin, Nucl. Phys. B211 (1983)
93; T.Appelquist, T. Takeuchi, M. Einhorn and L.C.R.
Wijewardhana,
Phys. Lett. B 232 (1989) 211;
A. Dobado and M.J. Herrero, Phys.Lett. B228 (1989) 495;
J.F. Donoghue and C.Ramirez, Phys. Lett. B234 (1990) 361;
B.Holdom and J.Terning, Phys. Lett. B247 (1990) 88;
A.Dobado and M.J.Herrero and J.Terron, Z.Phys. C50 (1991) 205, 465;
S.Dawson and G.Valencia, Nucl.Phys. B352 (1991) 27;
M.Golden and L.Randall, Nucl. Phys. B361 (1991) 3;
D. Espriu and M.J. Herrero, Nucl. Phys. B 373 (1992) 117;
T. Appelquist and G. Triantaphyllou, Phys. Lett. B278 (1992) 345;
J.Bagger, S.Dawson and G.Valencia, Fermilab-Pub-92/75-T, (1992);
T. Appelquist and G-H. Wu, Phys.Rev. D48 (1993) 3235}

\item{[30]}{ M.J.Herrero and E.R.Morales,
FTUAM 9411, Oct 1994; A.Nyffler and A.Schenk HUTP-94/A012 (1994)}

\vfill
\eject

\beginsection{\bf Figure Captions}

\bigskip\bigskip
\item{\bf Fig. 1.-} {Tree level diagrams contributing to the
scattering amplitude $A(W_{L}^{+} W_{L}^{-} \to W_{L}^{+} W_{L}^{-})$
in the SM.}
\medskip
\item{\bf Fig. 2.- } {Tree level diagrams contributing to the Goldstone
scattering amplitude $A(\omega^{+} \omega^{-} \to \omega^{+}
\omega^{-})$ in the SM.}
\medskip
\item{\bf Fig. 3.- } {Tree level diagrams contributing to the first
next to leading correction $A({\tilde W}^{+} \omega^{-} \to \omega^{+}
\omega^{-})$. The
external gauge line is contracted not with a $\epsilon_{L}^{\mu}$ but
with a $v_{\mu}$. To get the complete $A({\tilde W}
\omega \omega \omega)$ one should consider all permutations.}
\medskip

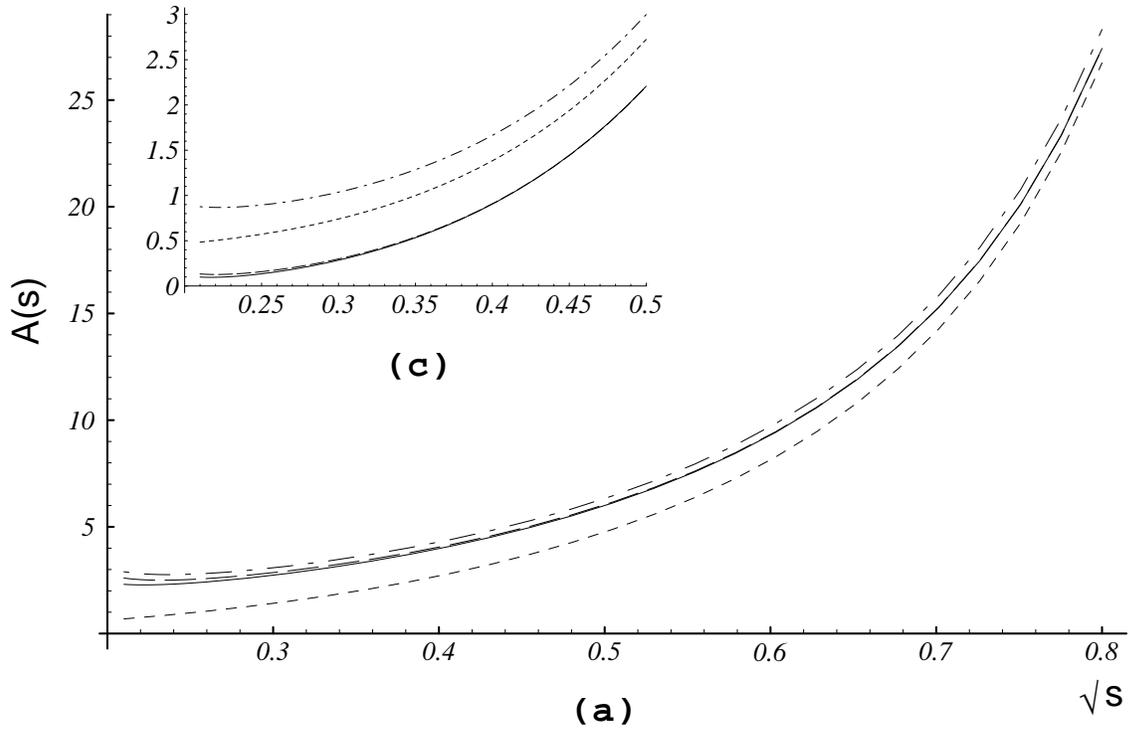
\item{\bf Fig. 4.- }{Comparison between the exact tree level
amplitude
of $W_{L}$'s in the SM (solid line) for three different angles
$\theta=\pi/4,\pi/16,3\pi/4$ ((a),(b),(c), respectively) and four
different approximations:
i) the standard one, i.e.,  $A(\omega^{+} \omega^{-} \to \omega^{+}
\omega^{-})$ with $g$ and $g^{\prime}$ set to zero (dashed line).
ii) The complete Goldstone
amplitude including $Z,\gamma$ interchange diagrams (dashed-dotted
line).
iii) The complete ${\cal
O}(g^{2})$ contribution, i.e. ii) plus the
contribution coming from the diagrams
(a) (d) of Fig.3 which are of ${\cal O}(g^{2})$
(long-dashed line, nearly invisible because overlaps the exact
result). iv) In addition we have plotted
in (b) an extra line (dotted) which
differs from iii)
in that all the denominators are expanded up to
${\cal O}(M^{2}/s,M^{2}/t)$ except for the
Higgs propagator structure that has been kept intact.
\medskip
\item{\bf Fig. 5.- } {Contribution of the ${\cal L}_{11}$
operator to the right hand side of the E.T. (leading and next to
leading amplitude).}
\medskip
\item{\bf Fig. 6.- } {Tree level diagrams contributing to the
scattering amplitude $A(W_{L}^{+} W_{L}^{-} \to W_{L}^{+} W_{L}^{-})$
in an Effective Chiral Lagrangian up to ${\cal O}(p^{4})$. The black
circle means that the vertex includes contributions from
${\cal O}(p^{2})$ and ${\cal O}(p^{4})$ operators
and the cross means that there are only contributions from
${\cal O}(p^{4})$ operators.}
\medskip
\item{\bf Fig. 7.- } {(a)-(e) are the tree level diagrams contributing
to the scattering amplitude $A(\pi^{+} \pi^{-} \to \pi^{+}
\pi^{-})$
in an Effective Chiral Lagrangian up to ${\cal O}(p^{4})$. (f) +
permutations are the first contributions coming from
the next to leading amplitude $A({\tilde W} \pi \pi \pi)$.}
\medskip
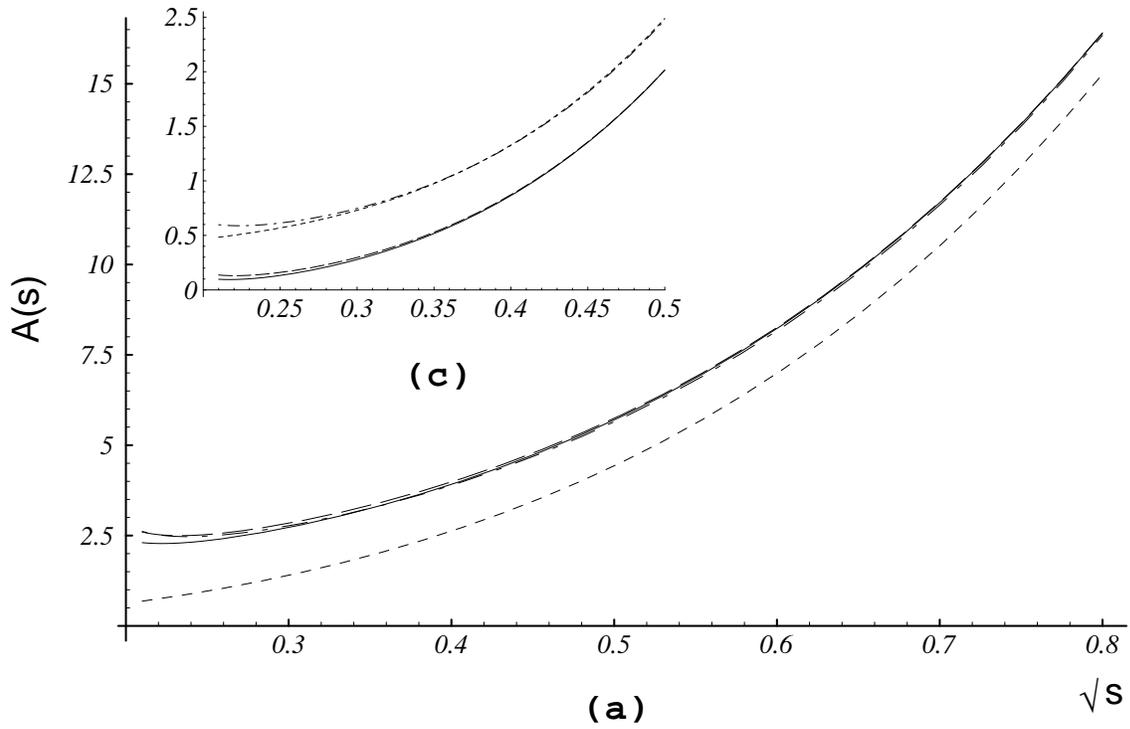
\item{\bf Fig. 8.- } {Comparison between the exact tree level
amplitude
of $W_{L}$ in an Effective Chiral Lagrangian for the particular
$a_{i}$'s that corresponds to the SM, i.e.
$a_{5}^{tree}=v^{2} / 8 M_{H}^{2}$ and the rest of
$a_{i}$ set to zero (same conventions for the lines as in
Fig.4).}
\medskip
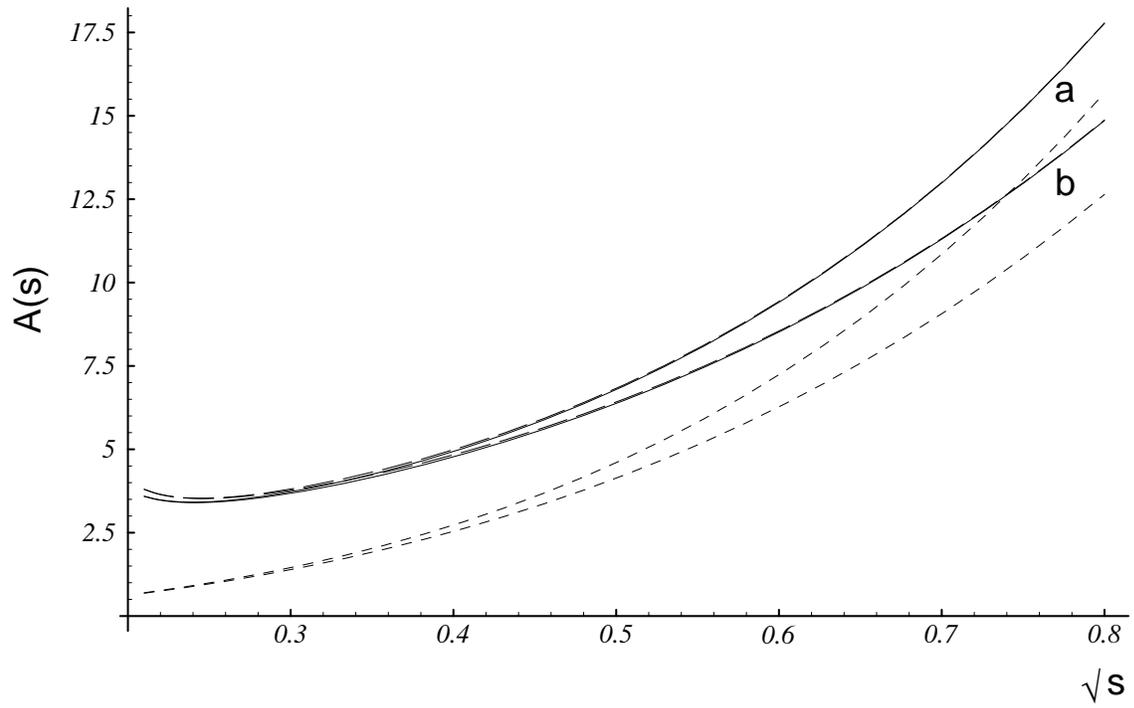
\item{\bf Fig. 9.- } {Comparison between the standard approach done in
the
literature $A(\pi^{+} \pi^{-} \to \pi^{+} \pi^{-})$ with $g=g^\prime=0$
(dashed line) in front of the exact result and our
approach (long-dashed line) for (a) $a_{5}=v^2/8
M_{H}^{2}$ and (b) $a_{5}= 16 / 384 \pi^{2}$. $\theta$ is $\pi/5$.}

\medskip
\vfill\eject

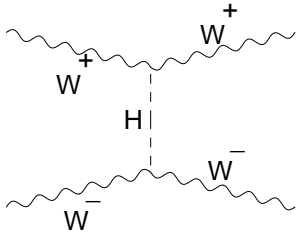
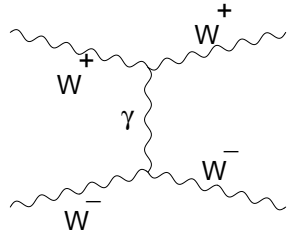
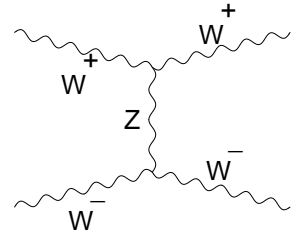
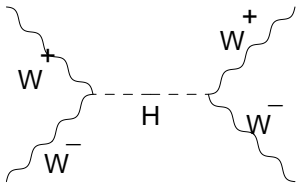
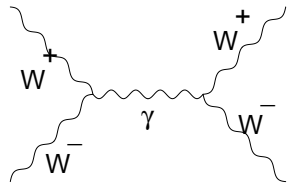
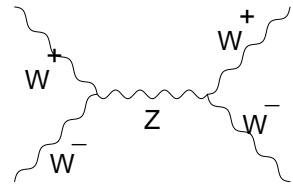
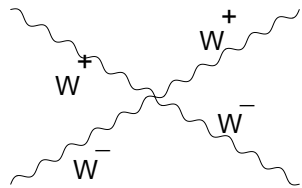
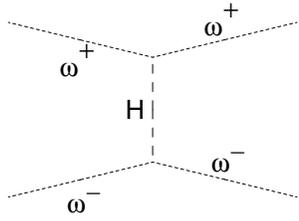
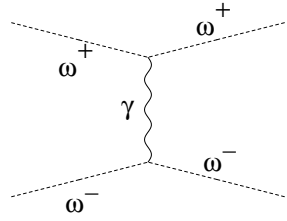
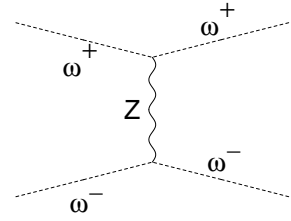
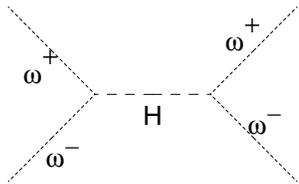
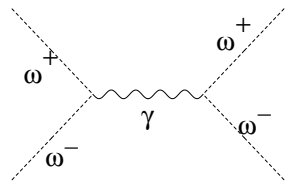
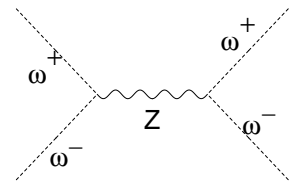
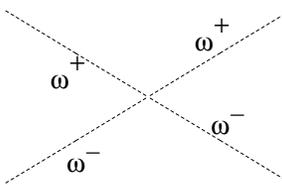
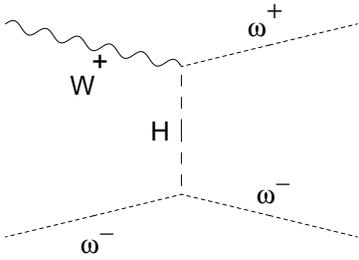
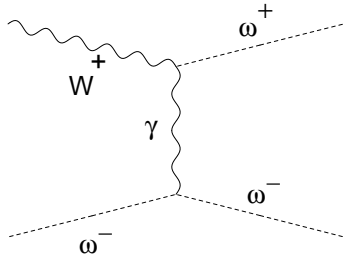
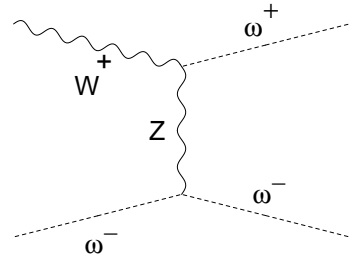
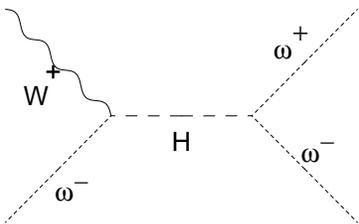
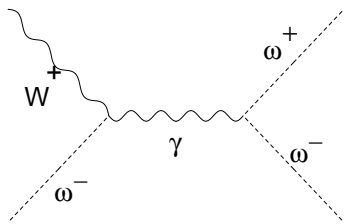
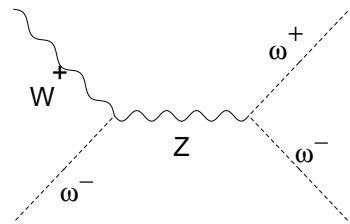
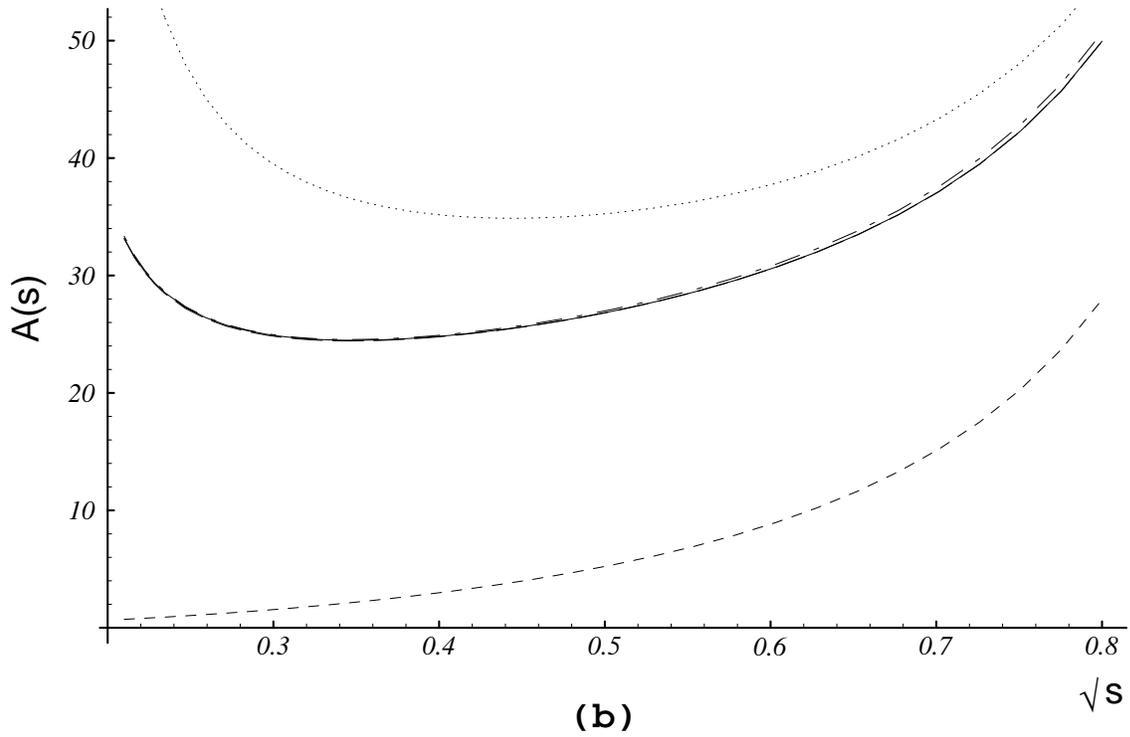
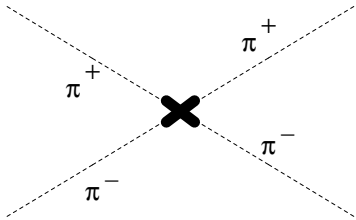
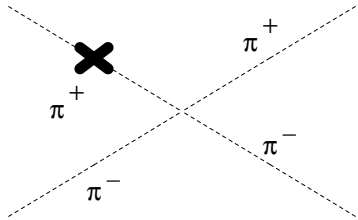
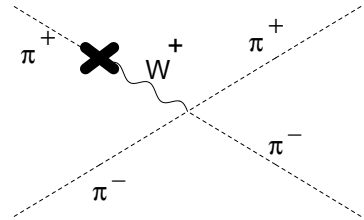
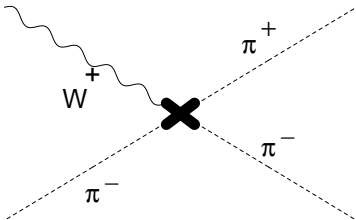
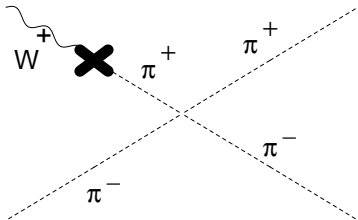
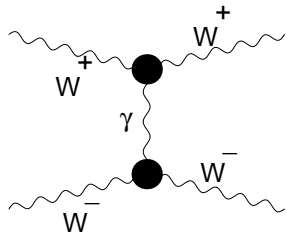
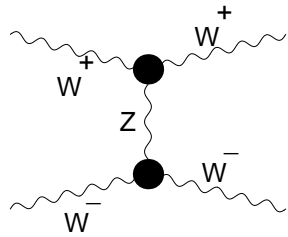
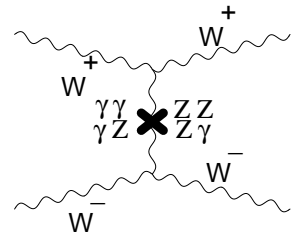
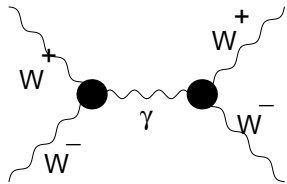
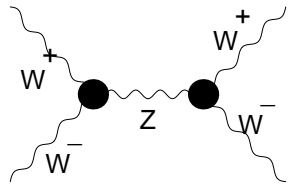
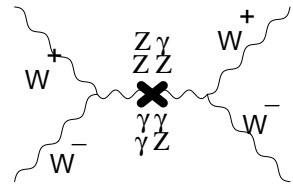
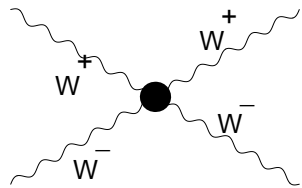
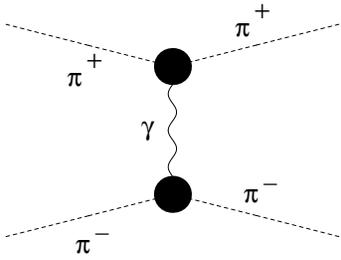
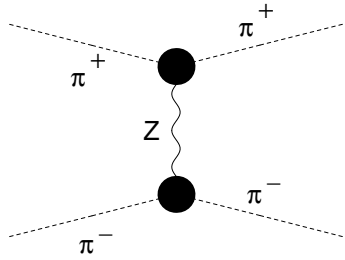
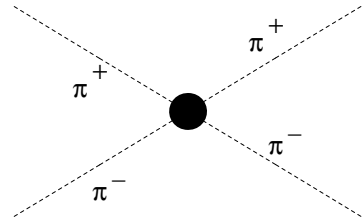
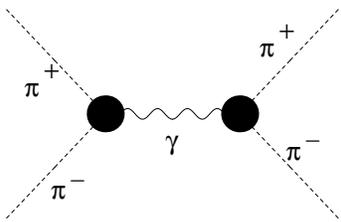
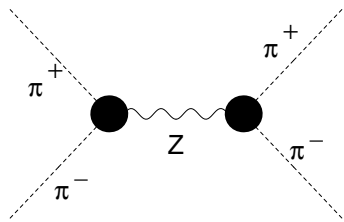
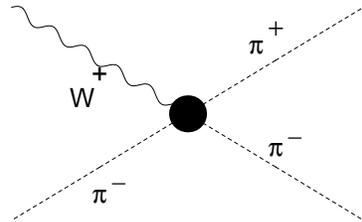
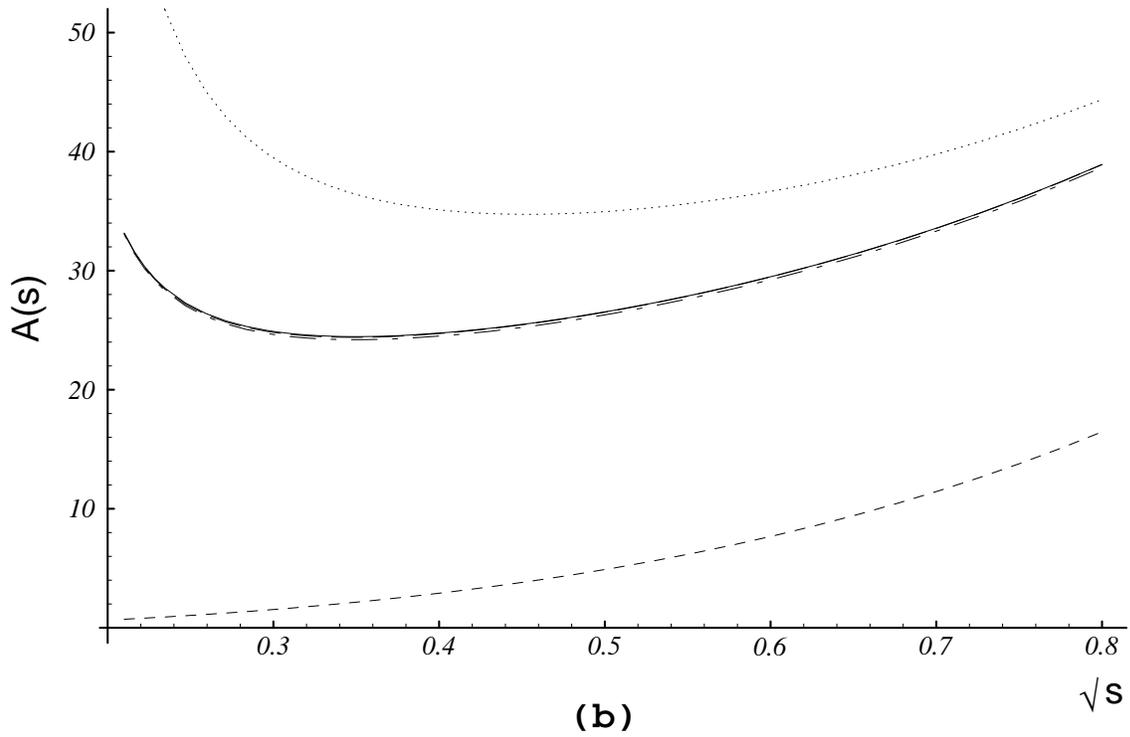
\bye